\documentclass[sigconf]{acmart} 
\usepackage{multirow}
\usepackage{pifont}
\usepackage{hyperref} 
\newcommand{\cmark}{\ding{51}}
\newcommand{\xmark}{\ding{55}}
\AtBeginDocument{
  }

\copyrightyear{2025}
\acmYear{2025}
\setcopyright{cc}
\setcctype{by-nc-nd}
\acmConference[FAccT '25]{The 2025 ACM Conference on Fairness, Accountability, and Transparency}{June 23--26, 2025}{Athens, Greece}
\acmBooktitle{The 2025 ACM Conference on Fairness, Accountability, and Transparency (FAccT '25), June 23--26, 2025, Athens, Greece}
\acmDOI{10.1145/3715275.3732037}
\acmISBN{979-8-4007-1482-5/2025/06}
\begin{document}

\title[Regulating Algorithmic Management]{Regulating Algorithmic Management: A Multi-Stakeholder Study of Challenges in Aligning Software and the Law for Workplace Scheduling}

\author{Jonathan Lynn}
\orcid{0000-0002-2682-0424}
\affiliation{
  \institution{The University of Texas at Austin}
  \department{School of Information}
  \city{Austin}
  \state{TX}
  \country{USA}
}
\email{jonathan.lynn@utexas.edu}

\author{Rachel Y. Kim}
\orcid{0009-0007-1483-4814}
\affiliation{
  \institution{Harvard University}
  \department{Department of Sociology}
  \city{Cambridge}
  \state{MA}
  \country{USA}
}
\email{rachelkim@g.harvard.edu}

\author{Sicun Gao}
\orcid{0000-0003-2524-4960}
\affiliation{
  \institution{University of California, San Diego}
  \department{Computer Science and Engineering}
  \city{San Diego}
  \state{CA}
  \country{USA}
}
\email{sicung@ucsd.edu}

\author{Daniel Schneider}
\orcid{0000-0001-6786-0302}
\affiliation{
  \institution{Harvard University}
  \department{Harvard Kennedy School}
  \city{Cambridge}
  \state{MA}
  \country{USA}
}
\email{dschneider@hks.harvard.edu}

\author{Sachin S. Pandya}
\orcid{0000-0001-7387-1307}
\affiliation{
  \institution{University of Connecticut}
  \department{School of Law}
  \city{Hartford}
  \state{CT}
  \country{USA}
}
\email{sachin.pandya@uconn.edu}

\author{Min Kyung Lee}
\orcid{0000-0002-2696-6546}
\affiliation{%
  \institution{The University of Texas at Austin}
  \department{School of Information}
  \city{Austin}
  \state{TX}
  \country{USA}
}
\email{minkyung.lee@austin.utexas.edu}

\renewcommand{\shortauthors}{Lynn et al.}

\newcommand{\jlynn}[1]{\textcolor{purple}{#1}}

\begin{abstract}
Algorithmic management (AM)'s impact on worker well-being has led to calls for regulation. However, little is known about the effectiveness and challenges in real-world AM regulation across the regulatory process---rule operationalization, software use, and enforcement. Our multi-stakeholder study addresses this gap within workplace scheduling, one of the few AM domains with implemented regulations. We interviewed 38 stakeholders across the regulatory process: regulators, defense attorneys, worker advocates, managers, and workers. Our findings suggest that the efficacy of AM regulation is influenced by: (i) institutional constraints that challenge efforts to encode law into AM software, (ii) on-the-ground use of AM software that shapes its ability to facilitate compliance, (iii) mismatches between software and regulatory contexts that hinder enforcement, and (iv) unique concerns that software introduces when used to regulate AM. These findings underscore the importance of a sociotechnical approach to AM regulation, which considers organizational and collaborative contexts alongside the inherent attributes of software. We offer future research directions and implications for technology policy and design.

\end{abstract}

\ccsdesc[500]{Human-centered computing~Empirical studies in HCI}
\ccsdesc[500]{Social and professional topics~Governmental regulations}

\keywords{Algorithmic Management, AI Regulation, Worker Management, Algorithmic Scheduling}

\maketitle
\vspace{-10pt}
\section{Introduction} 
Employer use of algorithms to hire \cite{raghavan2020mitigating, langenkamp2020hiring}, schedule \cite{kantor2014working, spektor2023designing}, surveil \cite{roose_machine_2019, goldstein_increase_2014}, and evaluate workers \cite{lee2015working, park2021hr}---known as \emph{algorithmic management} (AM) \cite{lee2015working}---has expanded beyond labor platforms and into traditional workplaces globally \cite{jarrahi2021algorithmic, park2021hr, lee2021participatory}. For example, in a 2024 OECD survey of managers across six countries, 74\% of them reported that their firms used software for managerial tasks \cite{milanez2025algorithmic}. Given the risks that AM poses to worker well-being \cite{zhang2022algorithmic, kantor2014working, park2021hr, awumey2024}, regulations emerged to control platform work (EU Platform Work Directive \cite{eu_directive_2024}) and employment and worker management (EU AI Act \cite{euai2024}), which encompasses algorithmic hiring (NYC Local Law 144 \cite{nyc_dcwp_nodate}) and scheduling (Fair Workweek Laws; Appendix \ref{sec:us-ordinances}). Some software developers have begun encoding such laws as product features to facilitate employer compliance \cite{7shifts_compliance_playbook, ukg_labor}. 

However, little is known about how real-world AM regulation unfolds across the entire AM regulatory process---rule operationalization, software use, and enforcement. The EU Platform Work Directive and EU AI Act are in their pre-implementation stage. Studies on implemented AM regulations for algorithmic hiring and scheduling focus on one part of the regulatory process \cite{groves2024auditing, petrucci2022persistent} or organizational practices \cite{haleylock2019evaluation, petrucci2022persistent, loustaunau2020combating}, often overlooking the alignment between software and law---that is, how well software design satisfies the desired outcomes of the law and, conversely, how compatible laws are with software development needs. Existing research on legally compliant software, often outside the AM context, focuses on challenges arising from legal ambiguity \cite{witt2023encoding, merigoux2022specification, otto2007addressing} and relies on laboratory settings without real-world usage data \cite{escher_code-ifying_2024, fan2020empirical, cejas2023nlpbasedcompliance}, or examines a specific regulatory stage such as enforcement \cite{giannoumis2015auditing, lassiter2024auditing}.

Our research addresses this gap by studying stakeholder experiences and challenges around the design and use of law-encoding software throughout the AM regulatory process in the context of workplace scheduling---one of the few examples of AM where an increasing use of software for managerial functions and the presence of laws have encouraged efforts to use software as a tool for compliance. Workplace scheduling refers to the practice in which managers allocate shifts and tasks to workers, a workflow management approach common in sectors like retail and fast food that employ hourly workers---who make up nearly 56\% of the U.S. workforce \cite{bls2023minimumwage}. While scheduling software is not new, recent machine-learning advances have increased work schedule uncertainty and its associated harms to shift workers \cite{kellogg2020algorithms, schneider2019consequences, uhde2020fairness, kantor2014working}. In response, starting in 2015, several cities and one State in the United States enacted so-called Fair Workweek (FWW) Laws to regulate work scheduling practices. As a result, many vendors now market their scheduling software as helping employers comply with these laws.

In this context, we conducted 38 semi-structured interviews with regulators, corporate defense attorneys, worker advocates, scheduling managers, and workers. We asked them about their experiences with scheduling software and FWW Law, including their beliefs about how such software affects FWW Law compliance and enforcement. Our findings (Figure \ref{fig:rulemaking} in Appendix \ref{app:findings-diagram}) suggest that the efficacy of AM regulation is shaped by: (i) institutional constraints that challenge efforts to encode law into AM software, (ii) on-the-ground use of AM software that affects its ability to facilitate compliance, (iii) mismatches between software and regulatory contexts that hinder enforcement, and (iv) unique concerns that software introduces when used to regulate AM.

These findings underscore the importance of a sociotechnical approach to AM regulation, one that considers organizational and collaborative contexts alongside the inherent attributes of software. To this end, we propose future research directions and offer implications for AM regulation as well as technology policy and design. We emphasize the necessity of multi-stakeholder engagement and boundary objects to foster collaboration throughout the regulatory process. We also suggest strategies for designing software interactions, establishing suitable levels of automation, gathering data that reflects the reliability of workplace decision-making, and improving regulatory data analysis capabilities, all aimed at better aligning software and the law. Our research contributes to human-computer interaction (HCI), software engineering, regulatory studies, and legal scholarship by providing empirical insights into challenges in regulating AM and developing compliant AM software.

\vspace{-5pt}

\section{Related Work}
Regulating AM is an area of growing interdisciplinary interest. We review prior work in regulating platform work, algorithmic hiring, and scheduling, and discuss research to develop software compliant with the law. We then situate our work within studies on technology in workplaces, worker-centered design, and HCI for policy.

\vspace{-9pt}
\subsection{Regulating Algorithmic Management} \label{sec:regulating-am} 
AM---the use of algorithms to automate managerial functions---pervades workplaces, promising to maximize corporate efficiency and profits \cite{jarrahi2021algorithmic, lee2015working, kellogg2020algorithms}. Mounting evidence of its harms to worker rights and well-being \cite{zhang2022algorithmic, zhang2023dataprobes, dzieza2020hard, us2024amazon} has prompted regulatory efforts across AM domains. However, empirical evaluation of AM regulation examining the role of software across the regulatory process is limited. For example, the EU AI Act includes provisions for regulating high-risk AI systems used for ``employment, worker management, and access to self-employment'' \cite{euai2024}. The EU Platform Work Directive \cite{eu_directive_2024} aims to regulate platforms to improve working conditions for gig workers. Yet, as they are still nascent, existing research largely consists of legal analysis of implementation concerns and the impact on the future of platform work \cite{veale2023fortifying, ponce2022regulating, aloisi2022gigging}, while lacking empirical evaluations of how rules will be enforced and how platforms will comply. In algorithmic hiring, existing regulations address bias in employment procedures, emphasizing measures of adverse impact \cite{uniformguidelines1978} and software audits \cite{nyc_dcwp_nodate}. To evaluate compliance with these laws, prior work analyzed hiring software companies' product websites \cite{raghavan2020mitigating}, self-audits \cite{wilson2021building, kassir_ai_2023}, job postings \cite{wright2024null}, and interviewed software auditors to understand their experiences \cite{groves2024auditing}. However, these efforts do not fully capture the state and challenges of implementing AM regulation, as they rely on the claims of employers or hiring software companies \cite{raghavan2020mitigating, wilson2021building, kassir_ai_2023, wright2024null} and the perspectives of auditors \cite{groves2024auditing}, whose work represents a single part of the AM regulatory process. 

In workplace scheduling, regulatory efforts led to FWW Laws and their integration into scheduling software \cite{7shifts_compliance_playbook, ukg_labor}. Research following these laws assessed employer compliance \cite{haleylock2019evaluation, harknett2021evaluating, harknett2021improving, harknett2021seattle} and their impact on worker well-being \cite{ananat2022effects, harknett2021improving} or schedule stability \cite{harknett2021evaluating, harknett2021improving, harknett2021seattle, haleylock2019evaluation}, with mixed results. However, these studies did not explore legal implementation and enforcement. Some research examined challenges regulators encountered while supporting employers with FWW adoption \cite{petrucci2022persistent, loustaunau2020combating}, yet lacked information on enforcement procedures and the perspectives of defense-side lawyers who interpret and implement laws for employers and software vendors. Most importantly, these studies have not examined the role of software throughout the AM regulatory process.

\vspace*{-12pt}

\subsection{Regulating Software} \label{sec:reg-software}
Regulating software systems has been a growing focus in software engineering research. However, prior work on encoding law into software remains largely theoretical or experimental, and research on real-world use lacks a holistic view of the full regulatory process. Studies on developing legally compliant software often focus on converting legal texts into software logic, highlighting how textual ambiguities complicate accurate translation \cite{witt2023encoding, merigoux2022specification, otto2007addressing}. Many argue that developers must collaborate with legal experts to align software with how they operationalize the law in real legal cases \cite{morison1992barrister, soltana2016model, boella2014critical, merigoux2022specification}, yet empirical studies on the effects of such collaborations \cite{witt2023encoding, escher_code-ifying_2024, digitalgovtnz_better_2022} are limited to experimental settings. For example, \citet{escher_code-ifying_2024} had computer science and law students translate excerpts of the U.S. Bankruptcy Code to software to simulate how software-legal teams would approach such a task, noting that their setup may not reflect real-world team dynamics.

Studies of real-world implementation often overlook actual software usage data or emphasize only enforcement or compliance perspectives. Researchers have examined software's compliance with the General Data Protection Regulation (GDPR) using the authors' own usage data \cite{fan2020empirical, guaman2021gdpr} and external video datasets \cite{shifa2020mulvis}, or examined privacy policy texts without validating their application in the software with empirical data \cite{cejas2023nlpbasedcompliance, rutledge2016privacyiot, bateni2022contentanalysis, torre2020privacy}. \citet{klymenko2022understanding} researched how corporate privacy experts applied GDPR's \textit{technical measures} for privacy, but examined only compliance actors' perspectives. \citet{lassiter2024auditing} investigated the practices of AI auditing professionals and \citet{giannoumis2015auditing} studied the challenges faced by third-party organizations conducting website accessibility certifications; yet they did not explore how regulators use these for legal action, nor include regulated companies' perspectives on being subject to their requirements. Real-world implementation of regulation hinges on how the regulators and the regulated interact, and focusing on one side risks missing critical aspects of the process.

\subsection{Technology for Workers and Policy} 
Research has analyzed how social contexts and organizational structures influence technology use, often deviating from design intentions \cite{norman1986user, kling1991cooperation, orlikowski2000using, bijker1995, christin2017algorithms}. For example, \citet{christin2017algorithms} highlighted how journalists and legal professionals used algorithms in ways that resisted and deviated from management's intentions and claims. Research also considers technology's influence on workers, such as \citet{fox2023patchwork}'s work observing the hidden labor incurred by essential workers during COVID-19 to address the breakdowns in the AI technologies, and \citet{park2021hr}'s examination of the impact of performance evaluations generated by HR software on worker well-being. Another area of study investigates the design of workplace technologies to prioritize worker needs. Participatory design research \cite{bjerknes_user_1995, bratteteig_disentangling_2012} has sought to involve and empower traditionally-excluded workers in AI design processes and balance the norms around decision-making \cite{spektor2023designing, kawakami2023sensing, hsieh2023codesigning, lee_webuildai_2019, zytko2022participatory}. For example, \citet{hsieh2023codesigning} engaged gig workers and policy-related stakeholders in workshops to generate ideas to improve well-being. Other studies designed shift scheduling algorithms to accommodate worker well-being \cite{lee2021participatory}, room-assignment algorithms with hotel staff \cite{spektor2023designing}, and eDiscovery tools with lawyers \cite{fernando2022uncommon}. Emerging research also tackles policy issues related to workplace technology design \cite{hsieh2023codesigning, hsieh2023designing, zhang2024data, lane2020regulatingplatform}, arguing for more collaboration between HCI and policy \cite{lazar2015future, yang2024future, junginger2013design, hochheiser2007hci, davis2012occupy} and offering recommendations and tools for policymakers \cite{hsieh2023designing, zhang2024data}. However, to our knowledge, there is limited empirical work on software that supports work duties \textit{and} compliance, particularly across the AM regulatory process.

\section{Methods}
\subsection{Participants} \label{sec:stakeholder-list}
We interviewed 38 participants from 5 stakeholder groups (Table \ref{tab:stakeholder-list} in Appendix \ref{app:demographics}). \textbf{Regulators} (n = 8) include staff at city-level agencies responsible for drafting, revising, and enforcing FWW Law (e.g., conducting investigations, negotiating settlements) as well as public education and outreach. We sampled staff from major U.S. cities with FWW Law who have engaged with scheduling software. \textbf{Corporate Defense Attorneys} (n = 6) advise and represent employers covered by FWW Law. They work with regulators and companies developing scheduling software (hereinafter software vendors) to discuss legal compliance issues. We prioritized attorneys with FWW Law and scheduling software experience, including two who also represent software vendors. \textbf{Worker Advocates} (n = 4) include employee-side attorneys, union organizers, and directors of worker advocacy groups. They represent workers in scheduling cases during enforcement (e.g., in court) and law-making stages (e.g., initiating legislation). \textbf{Scheduling Managers} (n = 7) include line and middle managers responsible for managing employee schedules using software, and \textbf{Workers} (n = 13) include cooks, cashiers, servers, and other staff in food service and retail industries covered by FWW Law. Regulators, advocates, and defense attorneys were recruited through purposive and snowball sampling. Managers and workers were recruited via sub-Reddit posts, Facebook advertisements, in-person visits to retail and fast-food chains, and snowball sampling, selecting those who used scheduling software. Participant demographics, work history, workplace details, and scheduling software usage were collected through a survey.

\vspace{-5pt}

\subsection{Interviews and Analysis} 
We conducted 60-minute semi-structured interviews between November 2022 and August 2023 via video conferencing, phone, or in person, based on participant preference. Workers and scheduling managers were compensated with \$50. Interview protocols asked participants to explain their role and experience with scheduling law and software, with prompts to adapt to each participant’s responses \cite{jimenez2021prompts}. Different protocols were used for each stakeholder role. For example, defense attorneys were asked about their experience litigating scheduling cases, and scheduling managers about their schedule creation process. (See Appendix \ref{app:interviewprotocol} for details).

Interviews were recorded and transcribed using Otter.ai with participant consent. One participant requested that their interview not be transcribed or quoted. We wrote memos after interviews to highlight emerging themes and coded interviews using DeDoose. Following \citet{deterding2021flexible}, we reviewed transcripts and notes to develop initial descriptive index codes, followed by a second round of analytic coding to identify themes, similarities, and differences across stakeholder groups. We discussed these during regular meetings. We also reviewed scheduling software by examining descriptions, FAQs, blogs, and supplementary documents on the public-facing websites of 38 products (details in Appendix \ref{sec:software-features-exhaustive}).

\vspace{-4pt}

\section{Workplace Scheduling}
Workplace scheduling is one of the few examples of AM where emerging laws and the increasing use of scheduling software for compliance make it a compelling case study for understanding how AM is regulated in the real world. This section summarizes relevant scheduling laws, software, and regulatory processes in the U.S.

\subsection{U.S. Laws to Regulate Workplace Scheduling} \label{sec:regulations}
Between 2015 to 2023, several U.S. cities and one state adopted Fair Workweek Laws (see Appendix \ref{sec:us-ordinances} for details) aimed at reducing precarious workplace scheduling practices. While FWW Laws vary by duties imposed, enforcement provisions, and coverage, they share key features. Employers must provide a \textbf{\textit{good faith estimate}} of regular schedules employees can expect for a given term of employment, including details such as weekly hours, day and time ranges, on-call shifts, and restrictions on schedule deviations. Employers must also give workers \textbf{\textit{advance notice}} (generally two weeks, posted in a conspicuous location) of actual schedules and maintain \textbf{\textit{records}} of schedules and changes for a minimum duration. If employers change schedules without \textit{advance notice}, they must pay a \textbf{\textit{premium pay}} to affected workers. Employers must offer open shifts to current employees before hiring new staff (\textbf{\textit{access to hours}}) and ensure a minimum 9- to 11-hour break between shifts on consecutive days (\textbf{\textit{right to rest}}). Such requirements also have \textbf{\textit{exceptions}}. For example, employers are exempt from \textit{premium pay} if workers \textbf{\textit{consent}} to schedule changes in writing. For a detailed summary of key provisions across jurisdictions, see Appendix \ref{sec:key-provisions}.

\subsection{Shift Scheduling Software as A Tool for Regulating AM} \label{sec:compliant-features}

Shift scheduling software promises to reduce labor costs, automate scheduling operations, and optimize workforce efficiency for employers \cite{lee2021participatory}. In our review of 38 product websites (details in Appendix \ref{sec:software-features-exhaustive}), we identified features that support and automate managers' scheduling workflows, such as staff demand forecasting, break allocations, shift recommendations, schedule violation alerts, and integration with enterprise software. We also found features designed for workers, such as shift swapping, direct and group messaging, and personal shift notes. While software can streamline scheduling processes, they have also contributed to an expansion of managerial control \cite{halpin2015subject}. Predictive analytics features allow managers to adjust staffing levels in response to sales data and customer demand, which can produce unstable schedules that introduce stress to workers’ lives, with implications for not only their own health and well-being \cite{schneider2019consequences, uhde2020fairness, bernstein2014manage, kantor2014working} but also that of their children and families \cite{carrillo2017instability}. A mother interviewed for a New York Times story \cite{kantor2014working} on the effects of Starbucks' use of scheduling software expressed feeling like her schedule controlled her life, from educational opportunities to childcare options.

With the passage of FWW Laws, many scheduling software are now marketed as helping employers comply with these laws to reduce unpredictable scheduling. Among the 38 software product websites we surveyed, 32 advertised features to aid compliance with applicable labor laws. Notably, 23 of these websites referenced specific FWW Law provisions. These software features include compliance alerts, record-keeping, and jurisdiction-specific customization \cite{deputy_FWW}. Although these websites did not disclose implementation details, if implemented correctly, scheduling software can significantly impact compliance and improve worker schedules.

\subsection{Rulemaking and Implementing FWW Law} \label{sec:rulemaking}
Enforcement agencies issue regulations and informal guidance documents to define how they intend to interpret and implement FWW Law. This process, known as \textit{rulemaking}, begins with an initial regulation proposed by the agency, followed by a notice-and-comment period to gather public feedback, and concludes with the release of the final regulation \cite{reginfo_gov_faq}. In some jurisdictions, stakeholders such as politicians, government agencies, worker advocacy groups, and corporate representatives participate \cite{nyccomments}, which can impact the resulting rules. To address compliance questions, regulators also provide informal guidance documents like fact sheets, FAQs, employer compliance tools \cite{dcwp_employer_tools}, as well as public outreach and training programs, workshops, or webinars. With the published regulations and informal guidance documents, software vendors decide what requirements the software must meet. Corporate legal and business teams use these documents to define internal compliance procedures and decide which software to use, which can influence workplace scheduling activities. Regulators use these documents to establish criteria for investigations and protocols for analyzing scheduling data to assess whether companies are compliant and determine any legal actions to pursue.

\section{Findings}
Our findings suggest that the efficacy of AM regulation is influenced by: (i) institutional constraints that challenge efforts to encode law into AM software, (ii) on-the-ground use of AM software that shapes its ability to facilitate compliance, (iii) mismatches between software and regulatory contexts that hinder enforcement, and (iv) unique concerns that software introduces when used to regulate AM. (Figure \ref{fig:rulemaking} in Appendix \ref{app:findings-diagram} provides an overview.)

\subsection{Institutional Constraints Can Challenge Efforts to Encode Law into AM Software} \label{findings:5.1}
Our findings suggest that regulatory guidance intended to clarify ambiguities in FWW Law may be insufficient for software development needs and hindered by adversarial dynamics during collaborative efforts between software vendors, defense attorneys, and regulators. Furthermore, a lack of sufficient financial incentives may also explain why jurisdiction-specific compliance requirements fail to get encoded into scheduling software.

\subsubsection{Regulatory Guidance is Insufficient for Operationalizing Rules into Software Requirements} \label{findings:5.1.1}
Despite regulators' efforts to provide guidance on FWW Law interpretation, defense attorneys expressed that rules were often too ambiguous to operationalize, especially in more nuanced workplace use cases. Regulatory guidance often included instructions for ``low-hanging fruit''---straightforward rules---but underspecified how ``ambiguous use case scenarios'' should be programmed (D3). For example, D1 worried if \textit{premium pay} penalties applied when workers asked to clock out early despite being paid for the full shift, which, though benefiting the worker, would be recorded as an employer-initiated schedule change. 

Without sufficient guidance, developers struggled to understand where new requirements should be encoded in the software. For instance, the ``Access-to-Hours'' provision mandates that employers offer shifts to current employees before making new hires. However, because hiring and scheduling are handled by different parts of the software, D3 explained that ambiguity over the application of this provision left him and software vendors uncertain about software implementation, thus requiring situated judgment calls:
\begin{quote}
    My understanding is software is, it's binary ... As a lawyer ... there isn't always a yes-no, zero or one answer. ... And that's a challenge that any software company is gonna have. (D1)
\end{quote}
Defense attorneys noted that software developers, therefore, relied on their own judgment or guidance from corporate business and legal teams (D3), increasing the risk that software behavior would diverge from regulators' intentions (D1). 

While this may, at its surface, suggest a need for detailed specifications tailored to more complex use case scenarios and software implementation needs, regulators expressed that some ambiguities were intentional to curb compliance tactics and software anti-patterns exploiting legal loopholes. While regulators emphasized a duty to assist employers in complying with FWW Law, they were cautious about their guidance being used to find workarounds. As a result, some chose not to explain everything in informal guidance documents (R3) or did so via private channels like email or phone (R6). Others refrained from publishing specific compliance information to preserve enforcement discretion (R2). The lack of guidance could also stem from their desire to avoid insinuating that companies would break the law. R3 described her agency's decision not to release its compliance guide as an effort to avoid the perverse outcome of providing instruction in ``here's how you don't break the law'', desiring it to be oriented around ``how to get to compliance'' instead. Finally, regulators pointed to how post-enactment rulemaking left room for loopholes (R6) and looser interpretations than intended due to pressures private interest groups are sometimes able to exert on the regulatory agency (R1).

\subsubsection{Barriers to Legal-Technical Collaboration Can Impede Software Development} \label{findings:5.1.2}
Regulators and defense attorneys highlighted challenges during the process of collaborating with software vendors that hindered their efforts to ensure FWW Law requirements were accurately applied to scheduling software. D2 and D3 noted that communication barriers between software vendors and legal experts contributed to vendors' misinterpretation of the law. D2 and R6 felt that vendors often failed to seek guidance proactively---such as during early rulemaking---and instead engaged with other stakeholders only after investigations uncovered issues in the code. R6 also observed a reliance on the wrong expertise and jurisdiction:
\begin{quote}
    They might rely on an accountant. The accountant might be a payroll person in a different department in another state who's advising about local scheduling laws. ... Because they're programming their payroll, they're not getting legal advice on the repercussions of the consent process. (R6)
\end{quote}

Furthermore, even when defense attorneys and vendors received clarifications, requirements sometimes got ``lost in translation'' as instructions passed from one group to the next, resulting in software being implemented incorrectly.
\begin{quote}
    ... you want to make sure that you’re not necessarily engaged in a game of telephone, where you’re saying one thing, it’s being interpreted differently. And that certainly is a very real problem ... as you go down from the legal chain and operation chain to the coding chain and back. (D3)
\end{quote}

\subsubsection{Software Vendors Lack Financial Incentives to Tailor Software to Local Laws}  \label{findings:5.1.3}
Regulators and defense attorneys noted that scheduling software frequently failed to meet the specific requirements of FWW Laws across jurisdictions. They believed that software vendors would not tailor software to local laws because of an insufficient business justification (D3, R8). R2 felt that the costs of losing business from employers subject to FWW Law ``wouldn't affect [vendors'] bottom line'', as they only apply to a few jurisdictions and industries. R8 felt jurisdiction-specific customizations would require major changes to how software vendors operate their business and have them ``lose an incredible amount of flexibility''.

Still, D4 suggested that there may be a financial incentive for software companies---especially those with a large market share in jurisdictions with FWW Law---to ensure that their products comply with local law, as doing so could offer a competitive advantage.

\begin{quote}
    And so there is one organization that has the tools to do that. ... They got ahead of the others. And that was their business benefit, I think. (D4)
\end{quote}
However, in jurisdictions with fewer employers, the financial incentive to comply may not be sufficiently compelling:
\begin{quote}
    But if, you know, let's say Hoboken were to come up with their own ... schedule change rules, like no software company is going to, you know, make something custom just for that. (R8)
\end{quote}

\subsection{On-the-Ground Use of AM Software Shapes Its Ability to Facilitate Compliance} \label{findings:5.2}
Participants shared that workplace operational demands and managerial practices shaped software usage patterns in ways that appeared to deviate from the intentions of FWW Law. Therefore, AM software's ability to help employers and workers comply with FWW Law and serve as evidence of compliance may be hampered. 

\subsubsection{Law-Encoded Software Is Not Designed to Meet Existing Workplace Routines and Demands} \label{findings:5.2.1}
Interviews with workers and managers suggest that software design sometimes clashes with workplace preferences and operational demands. As a result, software is either unused or used in ways that deviate from its design (e.g., consent collected after the shift instead of before). This creates ambiguities for regulators when analyzing scheduling records to accurately assess actual workplace scheduling activity and shapes how FWW Law is implemented. Many workers and managers preferred using other methods---enterprise applications (W1, W6, W9, M1, M7), phone calls and group chat platforms (W1, W4-6, W8, W9, W11, M1, M5), paper schedules (W12, W13, M5), or in-person processes---to communicate and manage shifts even when scheduling software provided the same features. For example, feeling it gave him priority for high-demand shifts, W3 would request new shifts in person:
\begin{quote}
    I decided to go to her office and told her like ``I've applied for that shift, can you like approve it?'' That's when she opened and I could see like 15 people had already applied before me, but since I was right there, already there, and the shift was in the next two hours, she approved mine. (W3)
\end{quote}

Defense attorneys (D1-3) felt that customer demands and busy operational flows at the store made it impossible for workers and managers to use the compliance software features. For example, if workers are scheduled for last-minute shifts, most FWW Laws require that worker consent for these changes be collected ``at or before the start of the shift'' (NYC Administrative Code 20-1221(d)), regardless of whether \textit{premium pay} is due. Defense attorneys felt that software features for collecting consent accordingly were not ``nimble enough'' (D3) to fit with the demanding work environment:
\begin{quote}
    And that's not how the fast food industry works. ... if you have a line, you're not gonna take someone off ... whatever they're doing to [log their consent]. That's just gonna delay things more. (D1)
\end{quote}
With FWW Law training typically done at the manager level (R5), it is their responsibility to ensure compliance. Yet R3 and D1 worried that this ``human element'' could cause breakdowns in how software supports compliance. For example, since workers may not have access to scheduling software while handling customers or demanding tasks, if they work overtime in the process, managers must interrupt their workflow to collect consent with the software (D1). D2 and D3 also worried that managers may be busy or absent during the day and forget to remind employees to initiate software consent procedures. According to D3, the problem of collecting consent was not technological, but operational.

\subsubsection{Data Generated by AM Software May Fail to Capture Noncompliant Managerial Practices} \label{findings:5.2.2}
Though curtailing coercive scheduling practices is one of the goals of FWW Law, regulators and defense attorneys still observed these practices, given managers' discretion in software use (D1, R3-5).
\begin{quote}
    And so we've seen systems where the manager can select the reason that the schedule change occurred. And then you know, a manager is selecting the reason that will result in premium not being paid. (R3)
\end{quote}
When managers can override or circumvent what the software allows, software procedures needed for compliance with FWW Law lose effect (R1, R6). R5 recounted examples of managerial behavior he considered to be ``the exact opposite of what [FWW] law envisions'' despite being compliant based on software records (R5). When managers were desperate for staffing, R5 felt that they used \textit{premium pay} ``as a carrot'' to encourage employees to stay late. He also observed managers telling workers not to count on the advance schedules as accurate forecasts of what they would actually be working, and argued that software constraints are needed to prevent such tampering from undermining the goals of FWW Law.

Managers' lack of knowledge of FWW Law (R1, M3) may explain some of this behavior. R1 argued that whether software is able to help managers achieve compliance is ultimately dependent on how well they know what the appropriate actions are under the law.
\begin{quote}
    Even if you have a perfect system that like is going to calculate it right, ... are the line-level managers being trained appropriately on like, what the law is, and putting in the scheduling changes appropriately? (R1)
\end{quote}
However, though managers felt that their scheduling decisions were dictated by inflexible staffing constraints (M2, M4, M6), advocates believed that this was also driven by managers' incentive to schedule according to the results employers want to see (A1) and to cut labor costs to boost their performance statistics and paycheck (A3).

\subsection{Mismatches between Software and Regulatory Contexts Hinder Enforcement}\label{findings:5.3}
Our findings highlight that a lack of concrete metrics and standards in scheduling rules, shortcomings in the quality of scheduling data, and insufficient agency data handling capacities may complicate the enforcement procedures and investigations regulators conduct.

\subsubsection{FWW Rules Lack Measurable Standards for Enforcement} \label{findings:5.3.1}
Software can aid regulators in enforcing FWW Law by easing data analysis and automating audits. However, this requires that regulators map legal requirements to explicit thresholds and calculations that the software can use to identify violations and determine the appropriate enforcement action. The goal of agency rulemaking is to determine these details, but some regulators felt that certain provisions still lacked ``teeth'' for enforcement after this process. For example, during investigations into whether employers provided \textit{good faith estimates}, R1 expressed that she ``never knew what to do with [the initial estimates of hours]'' received from employers, as the law did not specify thresholds for determining when deviations between those estimates and workers' actual schedules would constitute a violation. Without these measures, R3 reflected that it was also difficult to determine what legal obligations or charges to bring against employers: 
\begin{quote}
    [The Good-Faith Estimate] wasn't necessarily meaningful, and there were ... really no obligations that like flowed from it ... You know, it's not gonna like give workers any kind of predictability actually, because it's not enforceable, or it's just too slippery. (R3)
\end{quote}

Thus, using the software would still require ``deeply internal judgment calls'', which can be ``tricky'' (R1), especially if regulators are uncertain of what enforcement actions to pursue.

\subsubsection{Inconsistent Data Complicates FWW Investigations} \label{findings:5.3.2}
With scheduling activities recorded digitally in software, employers can produce the records regulators require to evaluate compliance with FWW Law more easily (D3). To obtain data for investigations, some regulatory agencies provide templates outlining how employers should report their data. However, the data received is not always in a structured format that can be readily synthesized (D2).
\begin{quote}
    Like we have a problem with one of the providers [who] did understand the law. But they had no ability to print up simple reports in the way the government wanted to see them. So even if you wanted to show compliance, it was like a jigsaw puzzle. (D2)
\end{quote}
Data was often scattered across separate systems, lacking integration and consistency, which made cross-referencing difficult (A3, D1, M7, R2, R8, W1). R4 and R8 noted that workers' schedules and timesheets often did not align. R6 reported that scheduling data used to claim exemptions frequently conflicted with payment types recorded in payroll. Typos, inconsistent worker names, and unstructured data also complicated the merging of data across systems (R8).
With these data issues, conducting investigations was cumbersome for regulators. R5 had to rely on sampling methods and estimates without sufficient data to make precise calculations or, like R1, manually sift through large volumes of data to find documentation to verify employer claims. R6 recounted repeatedly requesting employers to provide more data and in the right formats:
\begin{quote}
    We suffer. ... [The investigators] do suffer. And it's really me trying to figure out each individual case and then trying to find a pattern... (R6)
\end{quote}

\subsubsection{Lack of Procedural Evidence Undermines the Validity of AM Data as Proof of Compliance} \label{findings:5.3.3}
Our findings suggest that the data used to demonstrate employers' compliance with FWW Law often lacked evidence of the necessary procedural requirements. Certain provisions of FWW Law specify procedural requirements dictating how, for instance, consent for last-minute changes must be collected or how workers should be notified of schedule changes. For example, to be exempt from \textit{premium pay}, an employer must not only receive worker consent but also have a one-on-one conversation with the worker (Seattle Code tit. 14, §§ 14.22.045(b)). However, regulators mentioned that when employers claimed to comply with these provisions, evidence of satisfying the procedural requirements was often missing from employer scheduling data.
\begin{quote}
    ... the workers never say like, ``Oh, yeah, they sat down with me.'' ... But yet the app or the algorithm is still gonna try to utilize the exemption because somebody has signed their name next to a box that checks, you know, says no premium pay for that. (R5)
\end{quote}
Moreover, the data often lacked timestamps that could help regulators determine if the records were up-to-date (R5).

Additionally, A3 felt that employers used the lack of data as proof that there is no evidence of noncompliance. Without data on these procedural requirements, regulators must obtain ``backup documentation'' such as employee testimonies (R1) and text messages (R6) to verify employers' claims. However, regulators sometimes took this as an opportunity to ``cut to the chase'' on the grounds that employers lacked proof of compliance (R5).

\subsubsection{Insufficient Data Capabilities Limits Agencies’ Ability to Enforce the Law with Software} \label{findings:5.3.4}
Agency capacity is key to regulatory enforcement. Without sufficient technical infrastructure (A3, R6), protocols for data analysis (R6), data analysis staff, and staff training programs (A3, R6, R7), FWW cases may be prolonged (R1), settled with narrower terms (A3), prioritized based on settlement size (R5), or require advocacy from other organizations (R7). Scheduling software can expedite enforcement by automatically producing large amounts of scheduling data for analysis of violations. But in order to make enforcement decisions, agencies must be equipped to process and interpret this data (R1, R6). 

\begin{quote}
    And so, and also [city agency] has invested in data scientists, which I think you really need. So a huge part of implementing or enforcing these laws for us is that we have two people on staff that just do data, like really understand data. (A3)
\end{quote}

Although participants (A3, R2, R3, R4, R8) spoke of agencies with teams dedicated to data analysis, newer jurisdictions lacked the technical infrastructure, tooling, staffing, and technical protocols present at more established agencies (A3, R6). The technical support needed thus depends on the agency context.

\subsection{Concerns Introduced by Software When Used to Regulate AM} \label{findings:5.4}
\subsubsection{Potential Advantages of Software as Tools for AM Regulation} \label{findings:5.4.1}
Our interviews surface software's potential for facilitating compliance with the law. Some regulators and advocates expressed that the presence of software features automating legal requirements could expose managerial practices of tampering with the software (R1). This could enable advocates to dispel corporate narratives around the impossibility of compliance (A1) and prevent companies from lying about what is actually taking place at their stores, such as when managers edit time punches (R7).
\begin{quote}
    And so the scheduling software, I think, is an important way of helping council members (lawmakers) understand, how will this actually be implemented, and sort of refuting some business narratives around how onerous compliance is. (A1)
\end{quote}

Scheduling software can also give managers and workers increased visibility and control over schedule changes (R1). Employers can use these tools as evidence of their compliance efforts:
\begin{quote}
    I have been able, in my opinion, to reduce demands from the city and ultimate settlements based on showing the efforts that have been made [for employers] to use electronic software providers... (D2)\end{quote}

Still, participants informed us of the challenges when using software for AM regulation due to the limitations in the interactions afforded in the software interface and the opacity of the internal mechanisms by which these software systems guaranteed compliance. We elaborate on these challenges in the following sections.

\subsubsection{Interaction Design Patterns Can Undermine Software’s Reliability in Regulating AM} \label{findings:5.4.2}
Although software promises greater compliance with FWW Law, stakeholders felt that interactions afforded by the software interface challenged its reliability in regulating managerial practices. To regulators, the way certain interactions were designed, while appearing to meet FWW requirements, seemingly ``engineered'' (R6) outcomes that contradicted its intentions. For instance, under FWW Law, workers are owed \textit{premium pay} if they stay over 15 minutes late. However, while helping software vendors implement this feature, R6 observed that the developers made waiving \textit{premium pay} the default option. Specifically, the interface required workers to first explicitly decline to waive \textit{premium pay} before showing the option to accept it on the next screen, which effectively minimized the amount paid to workers.
\begin{quote}
    And [the software vendors] were like, ``[All the consent options are] there, but it's not shown to them initially. You have to click decline and then you would get here.'' I'm like, ``Well, that's cheating.''  (R6)
\end{quote}
R6 also noted that elements in the software interface asking for worker consent were often ``language-based''---written with dense and complex legal jargon. Workers with lower literacy would thus need to rely on managers to determine what button to select, which could result in them getting told to waive \textit{premium pay}, favoring the employer without reflecting what actually took place. Similarly, A1 worried that using software to satisfy procedural requirements---such as the mandated one-on-one conversations after a last-minute schedule change---could make the interaction less personal, eroding its value for workers and undermining the rule’s original intent.

Software interaction patterns also complicate defense attorneys' efforts to help employers demonstrate compliance. For example, employers can claim \textit{premium pay} exemptions if they can prove that last-minute schedule changes were initiated by the worker (e.g., clocking out early). To support this, a software vendor represented by D3 offered a dropdown menu for workers to indicate a reason when clocking out---such as (A) their own request or (B) their manager's. If the worker selected (A), the employer would be exempt from paying them \textit{premium pay}. However, according to D3, regulators argued that requiring workers to specify a reason was inherently coercive and unacceptable for compliance. Employers can also avoid violations if workers consent to the schedule change. However, software features for recording worker consent also created challenges for defense attorneys. D5 mentioned that features for collecting consent often relied on ``check[ing] a box'' instead of a ``hand-drawn'' signature. As a result, workers would often claim their consent was falsified, making it challenging for defense attorneys to prove that consent procedures were followed.

\begin{quote}
    And so we will often see plaintiffs allege there wasn’t---that they don’t recall [giving consent]. (D5)
\end{quote}

These examples illustrate that, when the design of a software user interface (UI) does not adequately consider the possible interaction and usage patterns that derive from its use, its ability to help regulators and employers with compliance can be hampered.

\enlargethispage{12pt}
\vspace{-3pt}

\subsubsection{Lack of Transparency and Variations Across Software Obscures How FWW Requirements are Implemented} \label{findings:5.4.3}
Software vendors often marketed their products as compliant with FWW Law (Section \ref{sec:compliant-features}). However, as implementation details are not available, regulators, defense attorneys, and managers expressed uncertainty and variation over whether and how compliance is handled.

First, despite falling under the jurisdiction of FWW Law, some software do not provide any support for managers to create or determine whether schedules are compliant with the law. For example, M7 mentioned that the work of ensuring compliance with FWW Law was a manual process:
\vspace{-2pt}
\begin{quote}
    There is no automatic updates or ties to any specific regulations. It is someone saying, ``Hey, new regulation'', and going in and manually updating, whether that be on a larger departmental scale or, ``Hey, ... we have to go through our schedule and make sure these ordinances are being followed''. (M7)
\end{quote}
This risked the rules not getting implemented. W2, who worked with M7, noted that their software allowed her colleagues to take shifts that violated their city's \textit{right to rest} requirement. 

In contrast, participants suggested that many software vendors ``overpromised'' (D2) that the schedules created by their software would automatically satisfy FWW Law requirements. For example, though M2 was led to believe that shifts would automatically be compliant with relevant laws, the software did not have mechanisms to detect compliance issues. Instead, M2 had to rely on separate reports and data processing to do so. In fact, M2 was skeptical of the state of compliance at their store and wished they had been audited by regulators. D2 and R6 reported cases where employers, despite relying on the software, found that they had actually violated FWW Law. R6 recounted an interaction with a family-owned franchise that had asked the corporate office for a software recommendation to help them with compliance. He recalled hearing the franchisee crying over the phone in disbelief, saying that in all their years in the business, they had never received so many violations.

Finally, some software handled compliance requirements by highlighting areas in worker schedules that violated scheduling laws, but required managers to decide how to handle the violations. For example, M6 mentioned that her software automatically checked if shifts were scheduled with a minimum 8-hour rest in between (\textit{right to rest}) and displayed a hazard symbol if violations existed. It even provided an explanation if she clicked the symbol. However, M6 felt it was still easy to create schedules violating this requirement:
\begin{quote}
    I wish it would actually prevent you from doing it. Like, I wish it would just [display]: ``Error. Not possible''. ...Because like, it's a very small triangle. And sometimes you overlook it. (M6)
\end{quote}
\enlargethispage{12pt}

This lack of transparency and variation makes it difficult to assess whether and how software is effectively regulating AM and adequately meeting the needs and expectations of all stakeholders.

\section{Discussion}
Beyond platform work \cite{lee2015working, zhang2022algorithmic}, AM is increasingly prevalent in traditional workplaces, covering diverse sectors such as public administration, finance and insurance, engineering, information and communication, real estate, education, social work, logistics, and healthcare \cite{milanez2025algorithmic, uhde2020fairness, ronnberg2010automating}. Based on our workplace scheduling case study, we propose future research directions and implications that can guide effective AM regulation.

\subsection{Sociotechnical Approach to Regulating AM}
Our research suggests that effectively regulating AM requires not only establishing legal rules and procedures governing work practices but also designing software that promotes the desired work practices and supports regulatory enforcement. This dual focus underscores the importance of a sociotechnical approach to AM regulation, one that considers institutional and collaborative challenges alongside the inherent attributes of software. These findings suggest that regulations focusing solely on technology development as products, such as the EU AI Act, may not adequately protect workers. In subsequent sections, we will discuss implications for supporting AM regulations and aligning software and the law.

Our study also raises new questions for regulatory studies and legal scholarship. Prior literature highlights the influential role of compliance professionals, such as corporate attorneys, in shaping how regulation is interpreted and implemented, frequently in ways aligned with organizational interests \cite{edelman2020working}. This \textit{managerialization} of law \cite{edelman2020working} may now be mediated by software, which already performs aspects of this role and affects the autonomy of these professionals. Future work can further investigate the impacts of software and how it should be designed to support this practice. Additionally, prior work argues that software can be a blunt instrument when used to enforce the law, interpreting it and configuring human behavior in a ``norm-enforcing'' manner \cite{diver2022digisprudence, goldoni2015politics, hildebrandt2008legal}. These works highlight its ``ruleishness,'' immediacy, and opacity \cite{diver2021interpreting, diver2022digisprudence}. However, our findings suggest that software’s role in regulating AM is not solely a function of these properties, and cannot be divorced from the social practices surrounding its use and the agency of ground-level users---managers and workers. Daily work practices shape how software is configured and used \cite{dourish2004context}, thereby influencing the meaning and implementation of the law. Further research should investigate the interplay between the two and examine how to develop \textit{norms} that accommodate worker priorities and workplace contexts. (Sections \ref{beyond}, \ref{multistakeholders}, \ref{interaction}, \ref{automation}, and \ref{validity} discuss related ideas.)

\subsection{Beyond Legal and Technical Expertise for Regulating AM} \label{beyond}
Our findings suggest that legal interpretation, technical implementation, corporate compliance, workplace demands, and regulatory enforcement all influence software regulating AM. Each involves distinct stakeholders, and eliciting their experiences and needs is crucial to developing software that effectively regulates AM. This suggests that only leveraging legal and technical expertise (Section \ref{sec:reg-software}) is insufficient for efforts to translate law into software. Determining the requirements to support these software affordances demands the expertise of regulators, managers, and workers. We encourage future legal translation work to consider the perspectives and expertise of these other stakeholders, whose needs are often overlooked in existing approaches. While we studied workplace scheduling, these implications are pertinent to other AM contexts with similar regulatory contexts and stakeholder involvement. A multi-stakeholder approach that explicitly examines the experiences of actors from every stage of the regulatory process is vital for designing policies and software that align with workplace needs, protect worker rights, encourage employer adoption, and support oversight. Furthermore, in so doing, we may be able to ensure that measures to regulate AM via software are legitimate \textit{ab initio} \cite{diver2022digisprudence}.

\vspace*{-6pt}

\subsection{Boundary Objects for Multi-Stakeholder Collaboration in AM Regulatory Process}\label{multistakeholders}
Our findings suggest that diverse stakeholders experience hindrances to collaboration, producing software that misses the requirements of the law. Prior work often argues that legal-software collaboration is difficult due to legal ambiguities and the epistemic divide between the two disciplines \cite{edelman2020working, escher_code-ifying_2024}. However, insufficient regulatory guidance may also be an obstacle to translating legal requirements into software (Section \ref{findings:5.1.1}). Additionally, regulators may leave requirements ambiguous to prevent loopholes (Section \ref{findings:5.1.1}), and engineers may misinterpret even well-defined instructions (Section \ref{findings:5.1.2}). Instructions can get lost in communication when traveling across teams and organizations (Section \ref{findings:5.1.2}), hindering collaboration and leading to misencoded rules. We argue that boundary objects---instruments that foster mutual understanding and knowledge exchange across diverse groups \cite{star1989boundary, star1999sorting}---are crucial for producing software as tools for compliance. To support developing software to regulate AM, boundary objects should foster a shared understanding of processes and needs across stakeholder groups, harmonize communication channels among distributed teams, provide process documentation to prevent miscommunication, and align stakeholder incentives by clarifying shared goals. 

Building on existing work in HCI (e.g., \cite{fernando2022uncommon, escher_code-ifying_2024}) and tools commonly used in software development, we offer some ideas for such boundary objects---personas, user stories, flowcharts, pseudocode, and data probes. Personas \cite{kaplan_personas_2022} and user stories \cite{rehkopf_atlassian_user_story} can help stakeholders in AM contexts build mutual understanding of each other’s needs. Asking stakeholders to create their own personas and user stories during collaborative sessions can clarify priorities and aid documentation for future discussions. Process flowcharts and pseudocode used to translate legal logic into code \cite{escher_code-ifying_2024} can also serve as boundary objects, but should be expanded to reflect workplace scenarios (Section \ref{findings:5.2.1}) and user interactions (Section \ref{findings:5.4.2}) that influence software’s ability to regulate AM. Illustrating stakeholder interactions, journeys, relationships, and service blueprints \cite{gibbons2017service_blueprint} in one unified diagram can give stakeholders involved in designing software a comprehensive understanding of the actors, resources, and workplace scenarios the system needs to support. Finally, data probes---which \citet{zhang2024data} used with workers and policymakers to visualize workers’ experiences using software platforms---can allow stakeholders to interactively explore use case scenarios collaboratively, establish shared awareness, and anchor conversations.

\subsection{Aligning Software and the Law}
\subsubsection{Designing Software Interactions that Safeguard Worker Rights}\label{interaction}
Our findings suggest that user interactions with scheduling software can produce outcomes that appear compliant but still harm workers. For example, even when software properly identifies when to collect consent, complex interaction patterns can undermine workers’ rights by ignoring the exploitative dynamics of worker-manager relationships (Section \ref{findings:5.2.2}) and nudge them toward choices that favor employers (Section \ref{findings:5.4.2})---practices known as dark patterns \cite{gray2018dark}. This is compounded by the fact that workers are often unaware of how their choices impact their rights and well-being (Section \ref{findings:5.4.2}) and that job demands may divert their attention from critical consent procedures (Section \ref{findings:5.2.1}). Therefore, designing worker-centric software interactions is critical to advancing the goals of AM regulation.

First, AM regulations should include strategies to detect dark patterns. Building on laws like the GDPR \cite{gdpr_text}, they can establish best practices for employers and software vendors to detect and prevent such practices in AM software. Enforcement must go beyond its traditional focus on static data to include analysis of software interfaces and user interactions. Research on detecting dark patterns in website consent banners \cite{gray2021, soe2020, nouwens2020, lu2024awareness} offers insightful guidance for such audits. Developers must also be attuned to the social contexts surrounding software use, as features intended to support users may appear mindless, coercive, or demeaning in certain settings \cite{alberts2024computers}. Empowering workers to contest manipulative design choices---such as through negotiation \cite{nguyen2020consent} or channels to communicate with regulators and developers \cite{alfrink2023contestable}---is essential. Additionally, allowing them to customize the UI and underlying algorithms---drawing on research in adaptive UIs \cite{shneiderman1997direct} and interactive machine learning \cite{amershi2014power}---can produce more responsive and context-sensitive software across varied workplaces and jurisdictions.

Second, workers must understand the legal and well-being implications of software use. Written notices, though limited  \cite{jordan2022strengths}, are a necessary starting point---but additional measures are needed. Efforts should be tailored to the user, particularly as certain populations may be more susceptible to dark patterns \cite{mildner2025comparative}. These should account for worker literacy levels to reduce reliance on managers for interpretation (Section \ref{findings:5.4.2}), such as by applying principles of semiotics to make the information more digestible \cite{desouza2005}. Ultimately, we advocate for \textit{process-oriented} approaches. Worker trainings are one strategy, but their effectiveness depends on \textit{who} delivers them and the \textit{attitudes} conveyed---both of which shape whether workers take them seriously. A potential remedy is to involve third parties, such as advocacy groups or regulators, to organize these trainings.

\vspace{-8pt}
\subsubsection{Deciding How Much Automation or Human Control to Allow} \label{automation}
In our findings, the degree of automation varied across software features designed for complying with AM laws, creating vastly different compliance experiences for those at the ground level (Section \ref{findings:5.4.3}). Some software automatically created schedules adhering to the relevant legal constraints, some identified potential violations but left final decision-making to managers, and others required managers' manual review. However, determining the ideal balance is an open problem, particularly for AM software that supports human decision-making that affects others. 

As a start, we highlight some key considerations. If implemented properly, full automation can efficiently prevent legal violations and avoid breakdowns caused by human involvement. If not, errors may go unnoticed and compound over time. Fully automated software may also not be adept at detecting nuances or contextual needs. Relying on complex algorithms may require data that is unavailable, is costly to obtain, or raises privacy concerns. Increasing human discretion in software processes offers flexibility to adapt to specific contexts and unforeseen scenarios, and gives humans greater control over decisions influencing their work. However, human discretion is also prone to error, coercion, and can be burdensome for decision-makers (Section \ref{findings:5.2.1}). Thus, creating transparent interactions \cite{lee_procedural_2019} that provide workers with relevant information---such as the potential consequences of their choices---and managers with compliance support through nudges or warnings, can promote better decision-making. A disclosure-based regulatory approach could further enhance accountability by requiring managers or firms to report their actions in alignment with regulatory goals \cite{fung2007full}.
\vspace{-10pt}
\subsubsection{Verifying the Procedural Validity of Evidence Used to Claim Compliance} \label{validity}
In AM, data used to substantiate compliance claims may be influenced by activities outside the software, challenging its validity as evidence. Our findings suggest that managers may override consent data to favor employers, pressure workers into last-minute shifts (Section \ref{findings:5.2.2}), and bypass procedural requirements with ``checkbox'' consent models (Section \ref{findings:5.3.3})---actions that undermine the reliability of AM software data as proof of compliance. Developing software to regulate AM requires capturing these additional data points so that regulators can accurately assess compliance, but identifying and recording all relevant actions---which can be problematic given the context \cite{roemmich2023values}---is an open problem and emerging research area in HCI and software engineering.

Design interventions can help reinforce the validity of compliance data. For instance, \citet{zytkofurlo2023onlinedating} suggest avoiding ``checkbox'' consent models, addressing barriers that marginalized stakeholders experience in consent processes, and initiating conversations around consent in the software. These strategies are relevant to AM as safeguarding and supporting workers' consent processes are key to ensuring that the consent data recorded in the software is valid. Software can exceed ``checkbox'' models by requiring additional interactions to verify that a worker is fully informed of their decision and mechanisms to withdraw consent should they change their mind \cite{nguyen2020consent, Schulenberg2023birdcage}. \textit{Data contextualization}, where users review their data and provide feedback on the claims made with it \cite{ortega2022menstrual, zheng2024trauma, razi2022instgram, badillourquiola2021teens}, can help workers elaborate on the interactions and contexts surrounding the moment their data was recorded and indicate directly to regulators whether the compliance claims based on this data are accurate. For example, they can then report that the recorded consent for waiving \textit{premium pay} occurred under fear of retribution from their manager. Regulators enforcing AM laws can provide anonymous channels for workers to submit this information to their agencies and to validate employers' claims.

\subsubsection{Supporting Regulatory Capacity to Implement and Audit Policy with Software}
HCI research can aid regulators in enforcing AM laws by addressing the challenges related to data quality (Sections \ref{findings:5.3.2} and \ref{findings:5.3.3}) and agencies' analysis capabilities (Section \ref{findings:5.3.4}). First, standardizing software record-keeping practices, data structures, and reporting requirements will be necessary. One approach is to adapt existing professional \cite{iso8000} and regulatory standards \cite{gdpr_text} for data management to the AM context. The growing volume of digital workplace data also requires regulators to have the tools, infrastructure, and human resources---especially technical expertise---to process and analyze it. These should be tailored to agency enforcement needs, considering factors like political climate, judicial decision-making, enforcement styles, and available resources. Enforcement standards will also need thresholds and calculations that can be readily integrated into AM software. In the FWW context, regulators highlighted a need for software to detect deviations between \textit{good faith estimates} and actual work schedules (Section \ref{findings:5.3.1}). Measuring these differences and the extent to which they pervade the workplace will require metrics, such as the fraction of schedules that deviate and the rate at which deviations for a worker occur, to determine when there is a sufficient legal case. Again, how these standards are quantified must also account for the dynamics of each individual regulatory context. 

\vspace{-7pt}

\section{Limitations and Future Work}
The open-ended and qualitative nature of interviews allowed us to elicit stakeholder experiences throughout the AM regulatory process and uncover challenges in AM regulation, an area relatively less understood. Future research can benefit from other methods such as surveys, observations, and archival data analyses. Regulators and defense attorneys who worked closely with or represented software vendors gave us insights into software vendors' perspectives, yet we did not directly interview software vendors and developers. Our study also took place in the U.S. in the context of workplace scheduling. Future research should involve software vendors and developers and explore other regulatory contexts.
\vspace{-7pt}
\section{Conclusion}
Given the growing societal impact of AM, regulatory efforts are emerging to govern AM systems. Drawing on a case study of workplace scheduling---one of the few AM domains with implemented regulations---our research identifies four key factors that shape regulatory efficacy and offers implications for policy and design. We highlight the importance of involving diverse stakeholders---regulators, employer- and software-side legal professionals, software developers, and end users---early in the regulatory process. Such engagement ensures that regulations are context-sensitive, responsive to worker needs, and practically enforceable. It also aids in distinguishing which requirements should be addressed through organizational practices versus encoded in software, and in making those requirements actionable for software development. Research on boundary objects that support cross-stakeholder collaboration and capture collaborative decisions will be critical. We further argue that AM regulation calls for new directions in software research---particularly in designing interactions and transparency mechanisms that uphold worker rights, determining appropriate levels of automation in compliance, collecting procedural data that reflects decision-making processes, and enhancing analytic capabilities for regulatory oversight. We hope these insights inform future research and regulatory practices that promote worker-centered AI in the workplace.

\section*{Researcher Positionality Statement}
The author team includes researchers at a variety of academic institutions covering diverse disciplines, including human-computer interaction, computer science, law, and sociology. We acknowledge that these professional affiliations and experiences influence the framing of our research questions, the design of our study, our access to and interactions with our respondents, and our data analysis. For instance, we acknowledge that our professional affiliations allowed us to use our personal networks in recruitment, which gave us unique access to hard-to-reach professional populations, such as corporate attorneys and regulators. At the same time, our work experiences differ greatly from those of service sector workers. The authors took great care to ensure that workers and managers who participated in this study were appropriately compensated for their time. We also de-identified their answers to the best of our ability. 

\begin{acks}
This work was supported by the following: National Science Foundation CCF-2217721, DGE-2125858, IIS-1939606 grants; Good Systems, a UT Austin Grand Challenge for developing responsible AI technologies\footnote{https://goodsystems.utexas.edu}; and the School of Information at UT Austin. We thank our participants for their valuable insights and trust in our work throughout the interview sessions. We are grateful to Alexander Boltz, Oterhime Eyekpegha, Meah Lin, and Esther Uzoma, who helped with interviews, data analysis, and diagrams; Apoorva Gondimalla, Woohyeuk (Kevin) Lee, Whitney Nelson, Angie Zhang, members of UT Austin's Doctoral Writing Seminar taught by Dr. Andrew Dillon, and the anonymous reviewers who provided invaluable feedback during the writing process.
\end{acks}

\bibliographystyle{ACM-Reference-Format}
\bibliography{references}

\clearpage
\appendix
\onecolumn
\section{Figure of Findings} \label{app:findings-diagram}
\begin{figure*}[h]
  \centering
  \includegraphics[width=0.965\textwidth]{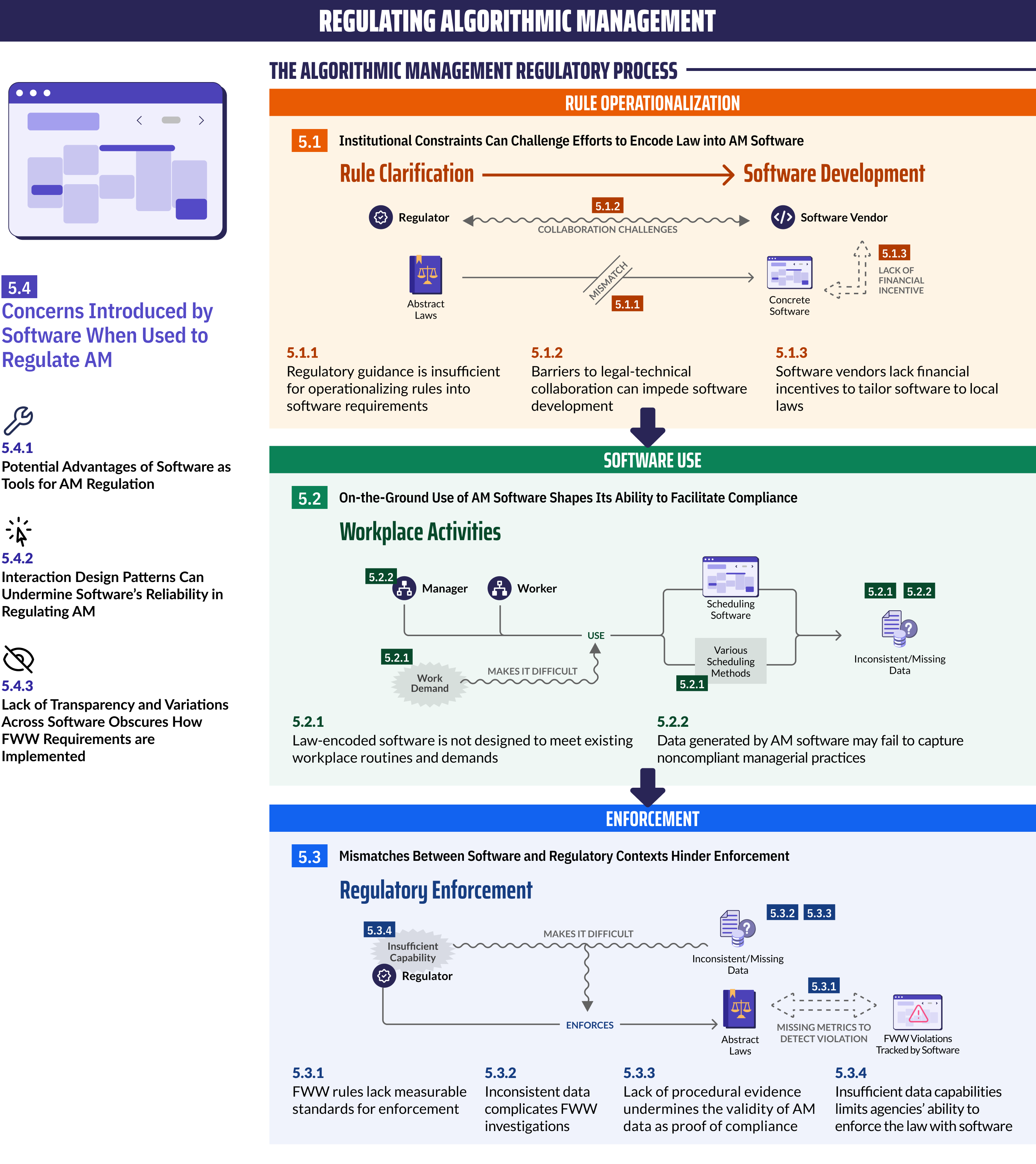}
  \caption{This figure illustrates the AM regulatory process, which encompasses the stages of rule operationalization, software use, and enforcement. It also highlights the challenges that stakeholders experience during this process and the unique concerns that software introduces as it is used for AM regulation. The numbers indicate the corresponding sections in our findings.}
  \Description{}
   \label{fig:rulemaking}
\end{figure*}\textit{\textbf{Step 1---Rule Operationalization}}: Once the enforcement agencies have issued rules and informal guidance documents following the rulemaking process, software vendors will use and interpret the requirements and instructions in these documents to determine the software specifications (e.g., UI features, software logic, data record-keeping protocols, calculations) needed so that employers using their products will be compliant with FWW Law. Employers also use these documents to determine which software products to purchase (e.g., evaluate which software is compliant with the relevant scheduling laws).

\textit{\textbf{Step 2---Software Use}}: After the software is developed and purchased by an employer, it is then given to managers and workers to be used in their daily work activities. By using this software, the understanding is that workers' and managers' work activities will be compliant with the law.

\textit{\textbf{Step 3---Enforcement}}: First, regulators (e.g., investigators) will use the rules and informal guidance to determine the measures and criteria they will use to instigate and carry out investigations. Specifically, the details in these documents will define the protocols regulators will use to analyze scheduling data and assess the degree of compliance and the corresponding enforcement action that follows. Then, once an investigation is initiated (e.g., via a worker complaint or class-action claim), regulators will request the data generated in \textit{Step 2} to evaluate whether employers are compliant with the law and determine whether legal actions need to be pursued if not.

\section{Fair Workweek Laws in the United States} \label{sec:us-ordinances}
\subsection{Scheduling Laws}
\begin{table*}[h!]
\centering
\begin{tabular}{llll}
\toprule
\textbf{Jurisdiction} & \textbf{State} & \textbf{Effective Date} & \textbf{Law Codified}                            \\ \midrule
Emeryville            & CA             & July 1, 2017            & Emeryville Mun. Code §§ 5-39.01 to 5-39.12       \\
Los Angeles           & CA             & April 1, 2023           & Los Angeles Mun. Code §§ 185.00 to 185.16        \\
San Francisco         & CA             & January 4, 2015         & San Francisco Police Code §§ 3300G.1 to 3300G.18 \\
Chicago               & IL             & July 1, 2020            & Chicago City Code §§ 6-110-010 to 6-110-170      \\
New York              & NY             & November 26, 2017       & N.Y.C. Admin. Code §§ 20-1201 to 20-1275         \\
Oregon                & OR             & July 1, 2018            & Or. Rev. Stat. §§ 653.412 to 653.490             \\
Philadelphia          & PA             & April 1, 2020           & Phila. Code tit. 9, §§ 9-4601 to 9-4611          \\
Euless*             & TX             & November 20, 2020       & Euless City Code §§ 95-1 to 95-6                 \\
Seattle               & WA             & July 1, 2017            & Seattle Code tit. 14, §§ 14.22.005 to 14.22.150  \\
\hline
\end{tabular}
\caption[]{Scheduling Laws in the United States in effect during Study Period. Jurisdictions not included: Berkeley, CA (effective Jan 1, 2023, fully operative Jan. 1, 2024) and Evanston, IL (effective Sept. 1, 2023). *Note: Euless City Code §§ 95-1 to 95-6 was repealed on Sep. 24, 2024.}
\label{tab:us-ordinances}
\end{table*}

\subsection{Key Provisions in Scheduling Laws across Jurisdictions} \label{sec:key-provisions}
\begin{table}[H]
\begin{tabular}{llccccccccc}
\toprule
\multicolumn{2}{l}{\textbf{FWW Law Provisions}}                       & \textbf{NYC} & \textbf{Chicago} & \textbf{Philadelphia} & \textbf{SF} & \textbf{Emeryville} & \textbf{LA} & \textbf{Seattle} & \textbf{Oregon} & \textbf{Euless}\\ \midrule
\multicolumn{2}{l}{\textbf{Good Faith Estimate}}    &  \cmark      & \cmark           & \cmark                & \cmark      & \cmark              & \cmark      & \cmark           & \cmark          & \cmark \\
\multicolumn{2}{l}{\multirow{2}{*}{\textbf{Advance Notice\textsuperscript{1} (Days)}}}        & Fast Food: 14     & \multirow{2}{*}{14}           & \multirow{2}{*}{14}                 & \multirow{2}{*}{14}       & \multirow{2}{*}{14}               & \multirow{2}{*}{14}       & \multirow{2}{*}{14}            & \multirow{2}{*}{14}           & \multirow{2}{*}{14}  \\
& & Retail: 3  & &&&&& \\
\multirow{2}{*}{\textbf{Right to Rest}}  & \textbf{Hours}       &  11      & 10           & 9                & \xmark      & 11              & 10      & 10           &  10         & \xmark \\
& \textbf{Penalty}       &  \$100      & 1.25$\times$ pay           & \$40                & \xmark      & 1.5$\times$ pay              & 1.5$\times$ pay      & 1.5$\times$ pay           &  1.5$\times$ pay         & \xmark \\
\multicolumn{2}{l}{\textbf{Access to Hours}}      &  \cmark      & \cmark           & \cmark                & \cmark      & \cmark              & \cmark      & \cmark           &  \xmark         & \xmark \\
\multicolumn{2}{l}{\textbf{Premium Pay}}     &  \cmark      & \cmark           & \cmark                & \cmark      & \cmark              & \cmark      & \cmark           & \cmark          & \cmark \\
\multicolumn{2}{l}{\textbf{Record Retention (Years)}}       &  3           & 3              &  2                  & 3             & 3                        & 3           & 3              &  3             & 1   \\
\multicolumn{2}{l}{\textbf{Right to Decline}} & \cmark      & \cmark           &  \cmark               & \xmark      & \cmark             & \cmark      & \cmark           &  \cmark           & \cmark   \\
\multicolumn{2}{l}{\textbf{Private Right of Action}} & \cmark      & \cmark           &  \cmark               & \xmark      & \cmark             & \cmark      & \cmark           &  \cmark           & \cmark   \\
\bottomrule
\end{tabular}
\caption{Overview of FWW Law provisions across jurisdictions. An (\xmark) indicates that the provision is not explicitly mentioned in the codified law of that jurisdiction. (1) This row shows the \textit{advance notice} requirement effective during the Study Period.}
\label{tab:fww-provisions-overview}
\end{table}

\subsubsection{Summary of FWW Law Provisions}
\begin{itemize}
    \item \textbf{Good Faith Estimate}: Employers must give workers an initial estimate of their expected work schedules.
    \item \textbf{Advance Notice}: Employees must be notified of their work schedules in advance. Employers are usually required to either post schedules in a conspicuous and accessible location in the workplace or send schedules directly to employees (usually electronically, so long as employees can access it on-site in the workplace).
    \item \textbf{Right to Rest}: Employees must be given a minimum duration of rest between their last shift on one day and first shift on the next.
    \item \textbf{Access to Hours}: Employers must give shifts to current employees before making new hires.
    \item \textbf{Premium Pay}: Also known as \textit{predictability pay}. The payments employers must give employees if schedule changes are not made with advance notice.
    \item \textbf{Record Retention}: Employers must retain records of schedules and scheduling activity for a minimum duration.
    \item \textbf{Right to Decline}: Employees have the right to decline schedule changes initiated by the employer without \textit{advance notice}.
    \item \textbf{Private Right of Action}: Employees can initiate civil action against employers for FWW violations.
\end{itemize}

\begin{table*}[h!]
\centering
\begin{tabular}{lccccccc}
\toprule
\multirow{2}{*}{\textbf{Jurisdiction}} & \multirow{2}{*}{\textbf{Deadline}}    & \multicolumn{4}{c}{\textbf{Required Information}} &           \multirow{2}{*}{\textbf{Threshold\textsuperscript{1}}}  & \multirow{2}{*}{\textbf{Length of Estimate}} \\
                             &     &  \textbf{Hours} & \textbf{Days and Times} &\textbf{Location} & \textbf{On-Call Shifts} &  &   \\ \midrule
\textbf{Emeryville}          &  First day of work  &   \cmark    & \xmark     & \xmark    & \xmark     & \xmark  & \makebox[2.7cm][c]{\xmark}     \\
\textbf{Los Angeles}         &  Time of hire       &  \cmark    &  \cmark     &  \cmark    &   \cmark   & \xmark  &  \makebox[2.7cm][c]{\xmark}      \\
\textbf{San Francisco}       &  Time of hire       &   \cmark    &  \cmark     &   \xmark   &  \cmark    &  \xmark &  \makebox[2.7cm][c]{1 month}    \\
\textbf{Chicago}             &  First day of work   & \cmark      &  \cmark     &  \xmark    &  \cmark    & \xmark  &  \makebox[2.7cm][c]{90 days}   \\
\textbf{New York}            &  First day of work   & \cmark      &   \cmark    &  \cmark    &  \cmark   & 15\%    & \makebox[2.7cm][c]{$\geq$7 days}    \\
\textbf{Oregon}              &  Time of hire        &  \cmark     &  \xmark     &  \xmark    &  \cmark    & \xmark  &  \makebox[2.7cm][c]{1 month}    \\
\textbf{Philadelphia}        &  First day of work   &  \cmark     &  \cmark     &  \xmark     &   \cmark   & \xmark  &  \makebox[2.7cm][c]{90 days}    \\
\textbf{Euless}              &  First day of work  &   \xmark    &   \xmark    &  \xmark    &  \xmark    & \xmark   &   Last date\textsuperscript{2}   \\
\textbf{Seattle}             &  Time of hire       &  \cmark     &   \xmark    &  \xmark    &  \cmark    & 30\%  & \makebox[2.7cm][c]{1 year}      \\
\hline
\end{tabular}
\caption[]{Details on Good Faith Estimate requirements across jurisdictions. An (\xmark) indicates that no explicit details are mentioned regarding that column in the codified law for that jurisdiction. (1) The threshold at which a schedule is considered to have deviated significantly from the Good Faith Estimate, measured by the percentage difference in hours between the Good Faith Estimate and actual hours worked. (2) The Good Faith Estimate in Euless must cover the period through the last date of the currently posted schedule.}
\label{tab:gfe}
\end{table*}

\begin{table*}[h!]
\centering
\begin{tabular}{lcccccc}
\hline
\multirow{2}{*}{\textbf{Jurisdiction}}  & \multicolumn{3}{c}{\textbf{No Change in Total Hours}}  & \multicolumn{3}{c}{\textbf{Adding Hours}}  \\
& \textbf{T$<$ 1}      & \textbf{1 $\leq$ T $<$ 7} & \textbf{7 $\leq$ T $<$ T*} & \textbf{T$<$ 1}      & \textbf{1 $\leq$ T $<$ 7} & \textbf{7 $\leq$ T $<$ T*} \\ \hline
\textbf{Emeryville}   & $W$  & $W$       & $W$   & $W$    & $W$        & $W$      \\
\textbf{Los Angeles}  & $W$  & $W$       & $W$    & $W$    & $W$        & $W$     \\
\multirow{2}{*}{\textbf{San Francisco}} & Case 1: 2 $\times$ $W$   & \multirow{2}{*}{$W$}              & \multirow{2}{*}{N/A}             & Case 1: 2 $\times$ $W$                   & \multirow{2}{*}{$W$}              & \multirow{2}{*}{N/A}                                               \\
                                        & Case 2: 4 $\times$ $W$  &                                 &                                  & Case 2: 4 $\times$ $W$ &                                 &                                   \\
\textbf{Chicago}                        & $W$                         & $W$                               & $W$                                & $W$                         & $W$                               & $W$                                    \\
\textbf{New York}                       & \$15                      & \$15                            & \$10                             & \$15                      & \$15                            & \$10                             \\
\textbf{Oregon}                         & $W$                         & $W$                               & $W$                                & $W$                         & $W$                               & $W$                                 \\
\textbf{Philadelphia}                   & $W$                         & $W$                               & $W$                                & $W$                         & $W$                               & $W$                                \\
\textbf{Euless}                         & N/A                       & N/A                             & N/A                              & N/A                       & N/A                             & N/A                               \\
\textbf{Seattle}                        & $W$                         & $W$                               & $W$                                & $W$                         & $W$                               & $W$                             \\   \hline
\end{tabular}
\caption[]{Details on Premium Pay requirements across jurisdictions (Part 1). The columns show the required premium pay for schedule changes that result in (A) no change in total hours and (B) additional hours. The intervals represent the amount of advance notice given to the worker of the schedule change, in days. For example, $T<1$ means that notice was given within 1 day of the scheduled shift. $T$* is the amount in days of advance notice required by each jurisdiction's FWW Law (e.g., 14 days). $W$ is the hourly wage of the employee who is receiving the last-minute schedule change. San Francisco Case 1: premium pay if the original shift is $\leq$ 4 hours. San Francisco Case 2: premium pay if the original shift is $>$ 4 hours.}
\label{tab:premiumpay1}
\end{table*}

\begin{table*}[h!]
\centering
\begin{tabular}{lccc}
\hline
\multirow{2}{*}{\textbf{Jurisdiction}} & \multicolumn{3}{c}{\textbf{Reducing Hours}}  \\
& \textbf{T$<$ 1}      & \textbf{1 $\leq$ T $<$ 7} & \textbf{7 $\leq$ T $<$ T*} \\ \hline
\textbf{Emeryville}   & min(4, $S_{old}$) x $W$    & $W$     & $W$    \\
\textbf{Los Angeles}  & $\frac{1}{2}\times W\times S_r$   & $\frac{1}{2}\times W\times S_r$  & $\frac{1}{2}\times W\times S_r$ \\
\multirow{2}{*}{\textbf{San Francisco}}  & Case 1: 2 $\times$ $W$                   & \multirow{2}{*}{$W$}              & \multirow{2}{*}{N/A}                                               \\
                                       & Case 2: 4 $\times$ $W$ &                                 &                                   \\
\textbf{Chicago}                        & $\geq \frac{1}{2}\times W\times S_r$                                          & $W$                                                                                                & $W$                                                                                                \\
\textbf{New York}                     & \$75                                                                        & \$45                                                                        & \$20                                                                         \\
\textbf{Oregon}                        & $\geq \frac{1}{2}\times W\times S_r$ &$\geq \frac{1}{2}\times W\times S_r$ & $\geq \frac{1}{2}\times W\times S_r$ \\
\textbf{Philadelphia}                   & $\geq \frac{1}{2}\times W\times S_r$ & $\geq \frac{1}{2}\times W\times S_r$ &$\geq \frac{1}{2}\times W\times S_r$ \\
\textbf{Euless}                        & $\frac{1}{2}\times W\times S_r$                                                 & $\frac{1}{2}\times W\times S_r$                                                   & $\frac{1}{2}\times W\times S_r$                                                  \\
\textbf{Seattle}                      & $\geq \frac{1}{2}\times W\times S_r$ & $\geq \frac{1}{2}\times W\times S_r$ & $\geq \frac{1}{2}\times W\times S_r$ \\ \hline
\end{tabular}
\caption[]{Details on Premium Pay requirements across jurisdictions (Part 2). The columns show the required premium pay for schedule changes that result in (C) reduction of hours (which includes canceling the shift altogether). The intervals represent the amount of advance notice given to the worker of the schedule change, in days. For example, $T<1$ means that notice was given within 1 day of the scheduled shift. $T$* is the amount in days of advance notice required by each jurisdiction's FWW Law (e.g., 14 days). $W$ is the hourly wage of the employee who is receiving the last-minute schedule change. $S_r$ is the length in hours reduced as a result of the schedule change. $S_{old}$ is the length in hours of the originally-scheduled shift. San Francisco Case 1: premium pay if the original shift is $\leq$ 4 hours. San Francisco Case 2: premium pay if the original shift is $>$ 4 hours.}
\label{tab:premiumpay2}
\end{table*}
\clearpage

\section{Participant Demographics} \label{app:demographics}
\begin{table*}[h!]
\centering
\begin{tabular}{rllll}
\toprule
\textbf{ID}  & \textbf{Stakeholder type}        & \textbf{Age} & \textbf{Gender}           & \textbf{Race}                    \\ \midrule
\textbf{A1}  & Advocate                         & 35-44        & Female                    & White                            \\ 
\textbf{A2}  & Advocate                         & 35-44        & Non-binary / Third gender & White                            \\  
\textbf{A3}  & Advocate & 35-44        & Female                    & White                            \\ 
\textbf{A4}  & Advocate                         & 35-44        & Female                    & Latino, Latina, or Latinx        \\ 
\textbf{R1}  & Regulator                        & 35-44        & Female                    & White                            \\  
\textbf{R2}  & Regulator                        & -            & -                         & -                                \\ 
\textbf{R3}  & Regulator                        & -            & -                         & -                                \\ 
\textbf{R4}  & Regulator                        & -            & -                         & -                                \\ 
\textbf{R5}  & Regulator                        & 35-44        & Male                      & White                            \\  
\textbf{R6}  & Regulator                        & -            & -                         & -                                \\ 
\textbf{R7}  & Regulator                        & 35-44        & Female                    & Hispanic or Latino/a             \\ 
\textbf{R8}  & Regulator                        & -            & -                         & -                                \\ 
\textbf{D1}  & Corporate Defense Attorney   & -            & -                         & -                                \\  
\textbf{D2}  & Corporate Defense Attorney   & -            & -                         & -                                \\  
\textbf{D3}  & Corporate Defense Attorney   & -            & -                         & -                                \\ 
\textbf{D4}  & Corporate Defense Attorney   & -            & -                         & -                                \\  
\textbf{D5}  & Corporate Defense Attorney   & -            & -                         & -                                \\  
\textbf{D6}  & Corporate Defense Attorney   & 45-54        & Male                      & White                            \\  
\textbf{W1}  & Worker                           & 18-24        & Female                    & Asian                            \\  
\textbf{W2}  & Worker                           & 25-34        & Female                    & Asian                            \\  
\textbf{W3}  & Worker                           & 25-34        & Male                      & Asian                            \\  
\textbf{W4}  & Worker                           & 18-24        & Female                    & White, Latino, Latina, or Latinx \\  
\textbf{W5}  & Worker                           & 18-24        & Female                    & White, Latino, Latina, or Latinx \\  
\textbf{W6}  & Worker                           & 45-54        & Female                    & White                            \\  
\textbf{W7}  & Worker                           & 18-24        & Male                      & Latino, Latina, or Latinx        \\  
\textbf{W8}  & Worker                           & 18-24        & Female                    & White                            \\  
\textbf{W9}  & Worker                           & 18-24        & Male                      & White                            \\  
\textbf{W10} & Worker                           & 45-54        & Female                    & White                            \\  
\textbf{W11} & Worker                           & 18-24        & Male                      & Latino, Latina, or Latinx        \\  
\textbf{W12} & Worker                           & 18-24        & Male                    & Latino, Latina, or Latinx                            \\  
\textbf{W13} & Worker                           & 35-44        & Male                      &  Black or African American      \\ 
\textbf{M1}  & Manager                          & 18-24        & Non-binary (Transgender)  & Asian                            \\  
\textbf{M2}  & Manager                          & 25-34        & Non-binary / Third gender & White                            \\  
\textbf{M3}  & Manager                          & 45-54        & Male                      & White                            \\  
\textbf{M4}  & Manager                          & 35-44        & Female                    & White                            \\  
\textbf{M5}  & Manager                          & 18-24        & Male                      & Black or African American        \\  
\textbf{M6}  & Manager                          & 18-24        & Female                    & White                            \\  
\textbf{M7}  & Manager                          & 35-44        & Male                      & White                            \\ \bottomrule
\end{tabular}
\caption{Overview of Participant Table: Participant IDs are prefixed with the letter signifying their stakeholder type. We also provide their age, gender, and race for additional context. Entries that are omitted were not collected to uphold privacy. Workers and managers completed the demographic survey before participation, while the others did so afterward.}
\label{tab:stakeholder-list}
\end{table*}
\twocolumn

\section{Scheduling Software Review} \label{sec:software-features-exhaustive}
\begin{table*}[h!]
\centering
\begin{tabular}{ll}
\toprule
\textbf{Software Product} & \textbf{Website URL}  \\ \hline
7shifts                   & \url{https://www.7shifts.com/labor-compliance/}                                                                           \\
ADP                       & \url{https://www.adp.com/what-we-offer/time-and-attendance/workforce-management.aspx}                                     \\
Agendrix                  & \url{https://www.agendrix.com/}                                                                                           \\
Altametrics               & \url{https://altametrics.com/labor-scheduling/}                                                                           \\
APS                       & \url{https://apspayroll.com/unified-solution/time-attendance/}                                                           \\
Buddy Punch               & \url{https://buddypunch.com/}                                                                                             \\
Clock Shark               & \url{https://www.clockshark.com/}                                                                                         \\
Clockify                  & \url{https://clockify.me/flsa-compliance}                                                                                 \\
ConnectTeam               & \url{https://connecteam.com/predictive-scheduling/}                                                                       \\
Crunchtime (Teamworx)     & \url{https://www.crunchtime.com/labor-and-scheduling/labor-law-compliance}                                                \\
Dayforce                  & \url{https://www.dayforce.com/how-we-help/dayforce/agile-workforce-management/scheduling\#faq-accordion}                                               \\
Deputy                    & \url{https://www.deputy.com/fair-workweek}                                                                                \\
Factorial                 & \url{https://factorialhr.com/}                                                                                            \\
Find my shift             & \url{https://www.findmyshift.com/}                                                                                        \\
Harri                     & \url{https://harri.com/compliance/}                                                                                       \\
Homebase                  & \url{https://joinhomebase.com/hr-compliance/labor-law-compliance/}                                                        \\
HotSchedules              & \url{https://www.fourth.com/solution/workforce-management-software/restaurant-labor-compliance-software}                  \\
Humanity                  & \url{https://tcpsoftware.com/products/humanity/compliance/}                                                               \\
Jolt                      & \url{https://www.jolt.com/}                                                                                               \\
Kronos (UKG)              & \url{https://www.ukg.com/resources/white-paper/guide-fair-work-week-laws}                                                 \\
Microsoft Teams           & \url{https://www.microsoft.com/en-us/microsoft-teams/staff-scheduling-shift-management}                                   \\
Open sim sim              & \url{https://opensimsim.com/the-fair-work-week-why-a-scheduling-tool-could-be-the-answer/}                                \\
Quickbooks                & \url{https://quickbooks.intuit.com/time-tracking/resources/predictive-scheduling-laws/}                                   \\
Replicon                  & \url{https://www.replicon.com/regulation/united-states-labor-laws/}                                                       \\
Sage                      & \url{https://www.sage.com/en-us/human-resources/employee-scheduling-software/}                                            \\
Schedule Base             & \url{https://www.schedulebase.com/index.html}                                                                             \\
Shiftboard                & \url{https://www.shiftboard.com/}                           \\
Sling                     & \url{https://getsling.com/blog/}                                                                            \\
SnapSchedule              & \url{https://www.snapschedule.com/}    \\
Trupay                    & \url{https://www.trupay.com/software-solutions/employee-scheduling}                                                       \\
When I Work               & \url{https://wheniwork.com/blog/fair-workweek}                                                                            \\
When to Work              & \url{https://whentowork.com/}                                                                                             \\
Work Axle                 & \url{https://www.workaxle.com/workforce-management/scheduling-and-rostering}                                              \\
Workday                   & \url{https://www.workday.com/en-us/products/workforce-management/scheduling.html}                                         \\
Workforce                 & \url{https://workforce.com/software/scheduling-software}                                                                  \\
Workjam                   & \url{https://www.workjam.com/} \\
Zoho                      & \url{https://www.zoho.com/}                                                                                        \\
Zoom Shift                & \url{https://www.zoomshift.com/}         \\  \bottomrule                                    
\end{tabular}%
\caption{These are the scheduling software products we reviewed, along with their corresponding product websites.}
\label{tab:software-all}
\end{table*}

\begin{table}[p]
\centering
\caption{Scheduling Software Participants Used. Retail Chain B1, B2, and B3 represent different stores of the same retail chain.}
\begin{tabular}{lll}
\toprule
\textbf{ID} & \textbf{Company}     & \textbf{Scheduling Software} \\ \hline
W1          & Local Café Franchise & Sling                        \\
W2          & Local Food Service A & Workday and Humanity         \\
W3          & Local Food Service A & Workday and Humanity         \\
W4          & Fast Food Chain A    & Teamworx                     \\
W5          & Fast Food Chain A    & Teamworx                     \\
W6          & Retail Chain A       & UKG Dimensions               \\
W7          & Retail Chain B1*     & MyTime (customized)          \\
W8          & Fast Food Chain B    & Teamworx                     \\
W9          & Retail Chain C       & Kronos (now UKG)             \\
W10         & Retail Chain D       & Dayforce                     \\
W11         & Retail Chain E       & In-house Software            \\
W12         & Retail Chain F       & WorkJam                      \\
W13         & Retail Chain B3*     & Workday                      \\
M1          & Local Café           & Homebase                     \\
M2          & Retail Chain B2*     & Kronos (now UKG)             \\
M3          & Local Food Service B & Homebase                     \\
M4          & Retail Chain E       & In-house Software            \\
M5          & Fast Food Chain B    & Workday                      \\
M6          & Retail Chain E       & In-house Software            \\
M7          & Local Food Service A & Workday and Humanity        \\\bottomrule
\end{tabular}
\label{tab:scheduling-software}
\end{table}
\subsection{Methods}
We compiled our scheduling software website list by first identifying all scheduling software products used and/or mentioned by our study participants (details in Table \ref{tab:scheduling-software}). We then supplemented this list by using Crunchbase's company search tool\footnote{https://www.crunchbase.com/} and gathering all software products with employee shift-scheduling features. We removed all products whose websites were not in English, did not provide sufficient information about their features, or did not market to the U.S., as FWW Laws would not apply to these products.

\subsection{Detailed Software Review Results}
\begin{figure*}[h]
  \centering
  \includegraphics[width=\textwidth]{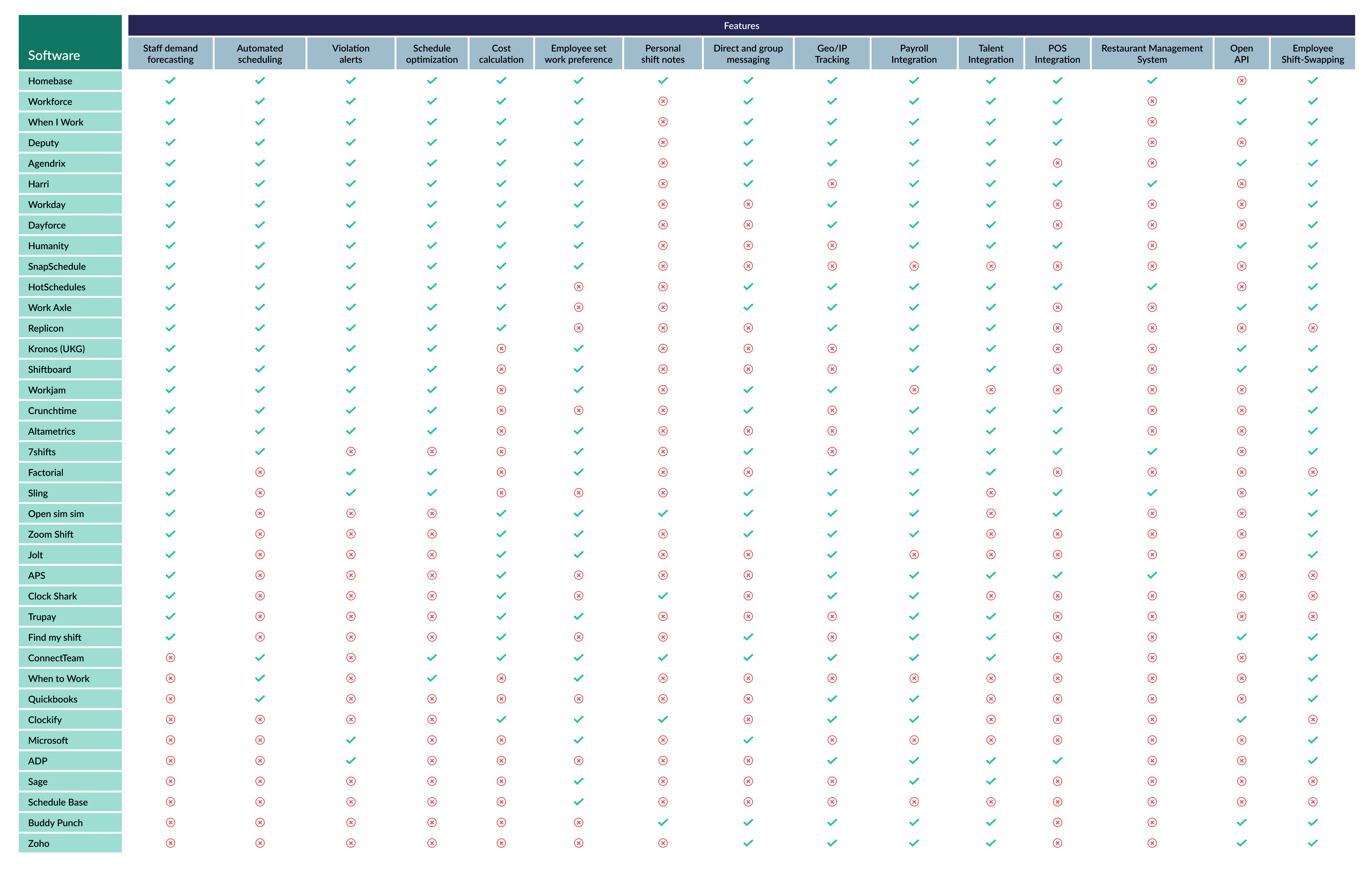}
  \caption{Overview of software features offered by different scheduling software companies ($N$=38)}
  \Description{}
   \label{fig:software_features_master}
\end{figure*}

Figure \ref{fig:software_features_master} shows the main features offered by all software products.

\subsubsection{Compliance Features}
\begin{figure}[h]
  \centering
  \includegraphics[width=\columnwidth]{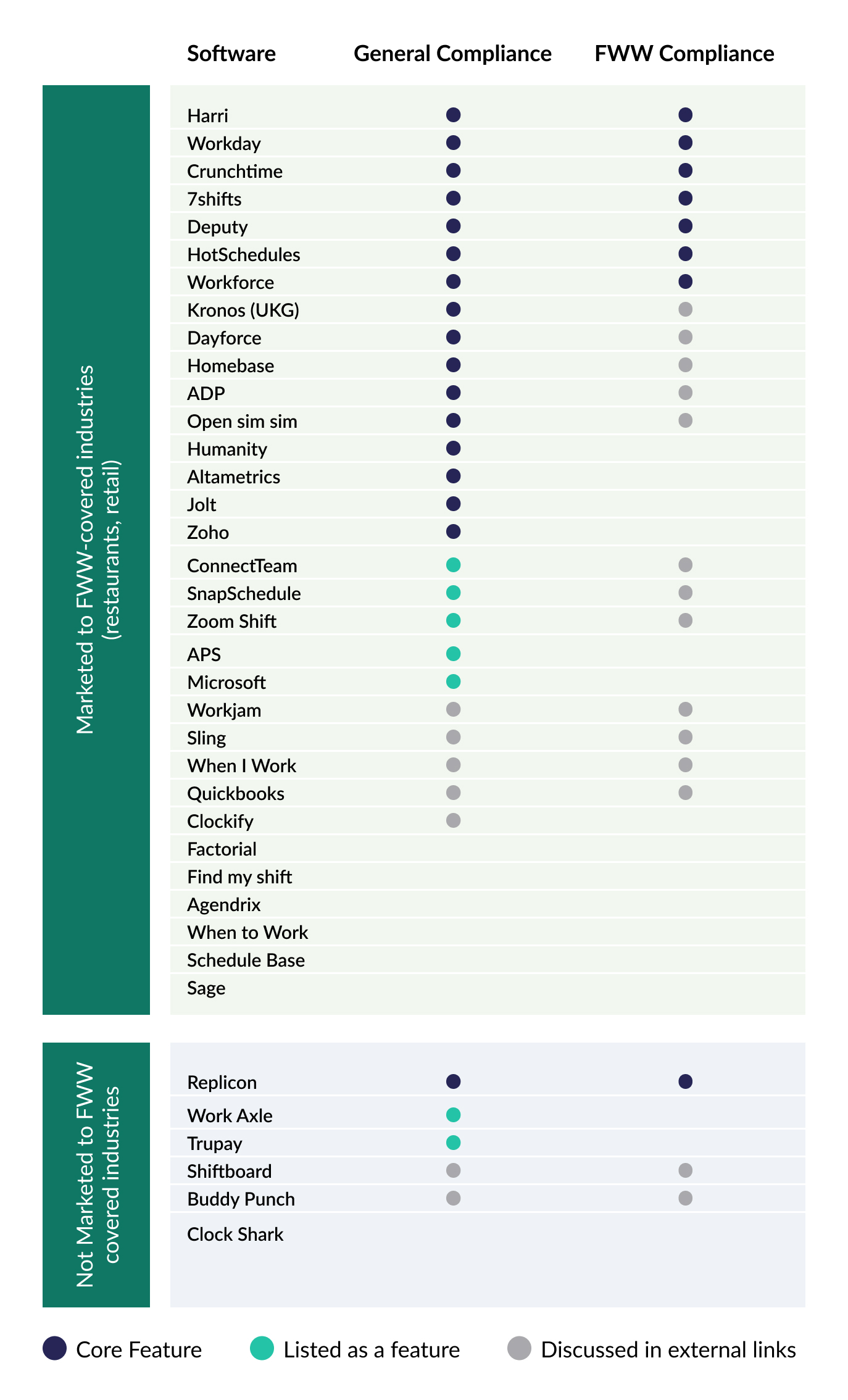}
  \caption{Overview of software features offered by different scheduling software companies ($N$=38)}
  \Description{}
   \label{fig:software_features_compliance}
\end{figure}

Companies advertised compliance in several ways. Some emphasized compliance as their main feature offering, such as in subtitles at the top of their websites or, in Shiftboard's case, explicitly providing compliance guarantees \cite{shiftboard}. Others did not explicitly mention compliance in their product descriptions. Instead, we only found information on how their products could help employers comply with relevant labor laws after searching through their FAQs, documentation, or blogs. For example, 7Shifts and Kronos (UKG) provided guidebooks for employers with background information about these laws and promises that their tools can help them with compliance. Kronos’ \cite{kronos_compliance_guide} guide featured a synopsis of the relevant laws and advised employer to be ``proactive with predictive scheduling compliance,'' which they used to motivate the case for using automated solutions such as their own. 7Shifts \cite{7shifts_compliance_playbook} provided more details on varying FWW requirements across different cities. They reported specific penalties, suggested ways to address common edge cases, and provided a compliance checklist for restaurant owners. Figure \ref{fig:software_features_compliance} gives an overview of how different software products advertised compliance. Table \ref{tab:software-all} documents all the websites we surveyed.

\clearpage
\section{Interview Protocols} \label{app:interviewprotocol}
We include the questions used in our stakeholder interviews below.
\subsection{Regulators}
    \textbf{Understanding the Stakeholder Role and Experience}
    \begin{itemize}
        \item How would you describe your job to someone who's never heard of it before?
        \item How has scheduling software fit into your prior or current work?
    \end{itemize}
    \textbf{Existing Regulations and Audit Process}
    \begin{itemize}
        \item I know that there are some existing regulations for AI scheduling tools. Could you talk us through [city]’s ordinance?
        \item What is your office’s role in this ordinance?
        \item What did the rulemaking process look like in [city]?
        \item How do these regulations guide your audit process?
    \end{itemize}
    \textbf{Typical Investigation Procedure}\\
    Walk me through your regular audit routine/workflow.
    \begin{itemize}
        \item How do you know and identify what to investigate?
        \item What triggers a review/audit?
        \item What are the data sources you use in the review? To what extent does the target employers’ scheduling software/timekeeping records play a role in any investigation or audit?
        \item How do you make the data and computation interpretable?
        \item How does this vary depending on the software a firm uses?
        \item How do you feel about the current process? How do you evaluate your process?
        \item Are there any tools or additional information you wish you could have to facilitate or improve your audit process?
    \end{itemize}
    \textbf{Past Investigation Experience}
    \begin{itemize}
        \item Can you tell us about an investigation when you looked at an AI scheduler?
        \item What are the components, features, or deliverables you examine within the software and/or require from the software provider in order to audit such a system? 
        \item What could have made your investigation easier?
        \item Can you think of any software features or components that can enable you to do things you weren’t able to do before?
    \end{itemize}
    
\subsection{Advocates}
    \textbf{Understanding the Stakeholder Role and Experience}
        \begin{itemize}
            \item How would you describe your job to someone who's never heard of it before?
            \item Can you tell us about your journey to becoming a worker advocate?
            \item What are the challenges you’ve faced and concerns you’ve observed from working with workers?
            \item What are the challenges you’ve faced and concerns you’ve observed from working with the managers/regulators/legal system/other actors?
        \end{itemize}
    \textbf{Policy/Organizing Involvement}
    \begin{itemize}
        \item I know that there are some existing policies governing workplace scheduling. Could you talk us through some of these?
        \item Thinking back over the campaigns for fair scheduling that you’ve worked on, can you focus on one, maybe Philadelphia or New York, and tell us more about that experience?
        \item What happened after the ordinance/law was passed?  Were you involved in rule-making?  What did the process look like?
        \item How did scheduling software feature, if at all, in the campaign, the crafting of the law, the debate around passage, and/or in rulemaking?
        \item In your opinion, what are some things that current policy on workplace scheduling software is doing well? 
        \item What are some areas in which policy around workplace scheduling software needs to be expanded/improved?
        \item To what extent does scheduling software hamper or facilitate compliance with the law?
        \item As an advocate, we know that you act as a broker in between parties with conflicting interests and goals. How would you describe the dynamics of the conversations about AI workplace scheduling software?
        \item What are the challenges you’ve faced from working with the managers/regulators/legal system/other actors?
    \end{itemize}
    \textbf{Legal Experience}\\
    If applicable, tell us about a legal case or brief you worked on that involved workplace scheduling.
    \begin{itemize}
        \item What was the nature of the problem/complaint between employer and employee?
        \item How did you come to learn about this case? What drew you to it?
        \item What were the various steps throughout your process?
        \item Tell us about how you did research/investigations to build your case.
        \item What external stakeholders did you involve for this case?
        \item What was the outcome?
        \item Did the dispute involve AI software?
        \begin{itemize}
            \item If so, how did that influence the development of the case?
            \item To what extent did software companies that make AI scheduling software become involved in these legal processes?
        \end{itemize}  
    \end{itemize}

\subsection{Scheduling Managers}
    \textbf{Questions for Managers}
    \begin{itemize}
        \item Tell me more about your job’s roles and responsibilities.
        \item How has scheduling software fit into your prior or current work?
    \end{itemize}
    \textbf{Experience with Scheduling}
    \begin{itemize}
        \item Describe your typical process of creating schedules.
        \begin{itemize}
            \item What are your priorities when creating schedules for the workers you manage?
            \item What are your concerns when creating schedules for the workers you manage?
            \item How long does it take you to create a schedule? 
            \item Do you need to get your schedules approved by anyone?
        \end{itemize}
        \item What are some problems you face when creating schedules for the workers you manage? 
        \item Do you use scheduling software or make them manually? 
        \begin{itemize}
            \item If software:
                \begin{itemize}
                    \item What software(s) do you use, and what do you use them for (if there are multiple)? Can you show me?
                    \item What aspects or features of the software did you find helpful?
                    \item What aspects or features of the software did you have concerns with?
                    \item Do you do anything manually in addition to the software?
                \end{itemize}
            \item If completely manual:
                \begin{itemize}
                    \item What parts of scheduling do you find easy?
                    \item What parts of scheduling do you find difficult?
                \end{itemize}
        \end{itemize}
        \item Are you familiar with the regulations and ordinances around scheduling (i.e. fair work week, predictability pay)?
        \begin{itemize}
            \item Do you feel like these scheduling regulations and ordinances are fair or unfair? 
            \item How do they affect the way you do your job?
        \end{itemize}
        \item Have you ever experienced an audit related to scheduling from your local Office of Labor Standards? 
    \end{itemize}

\subsection{Workers}
    \textbf{Questions for Workers}
    \begin{itemize}
        \item Can you tell me about your experience working at [company] so far?
    \end{itemize}
    \textbf{Experience with Scheduling}
    \begin{itemize}
        \item Can you walk me through the process of how your work schedule gets made?
        \item Do you usually like your schedule? 
        \item Are you familiar with your manager’s or company’s scheduling policy? How do you feel about it?
    \end{itemize}
    \textbf{Experience with Workplace Software}
    \begin{itemize}
        \item Could you walk me through your scheduling software? (Would you be able to show me?)
        \item What are your thoughts on the scheduling software you use at work?
        \item Does the software have ways that let you specify your preferences for shifts? Limit or update your availability before the schedule comes out?
        \item Does the software let you request or make changes to your work schedule once it is published?
        \item Can you swap shifts with co-workers?  How does that work?
        \item Has your manager ever changed your work schedule on the software? What happened? How did that play out?
        \item Is there anything else about your scheduling software that would be good for us to know? 
    \end{itemize}
    \textbf{Familiarity with Scheduling Ordinances}
    \begin{itemize}
        \item Are you aware that there are regulations around workplace scheduling (predictability pay, etc)? Are you familiar with any ordinances in particular or some of the details? 
            \begin{itemize}
                \item Do you understand the jurisdictions and scenarios that these ordinances apply to?
            \end{itemize}
        \item Do you know what rights you are entitled to under these ordinances?
    \end{itemize}
    If none, introduce the stakeholder to some relevant regulations in their city.
    \begin{itemize}
        \item Have you seen any of these adopted or put into practice in your workplace? 
        \begin{itemize}
            \item If you use scheduling software in your workplace, have you seen these regulations integrated into the tools?
        \end{itemize}
        \item How do you feel these regulations represent and uphold your preferences, needs, or values?
        \begin{itemize}
            \item If they are integrated into software, how well or poorly do you think the scheduling software interprets/applies these regulations and upholds your preferences, needs, or values?
        \end{itemize}
    \end{itemize}

\subsection{Defense Attorneys}
    \textbf{Questions for Defense Attorneys}
    \begin{itemize}
        \item How would you describe your job?
        \item How often do you encounter issues involving worker scheduling in your legal work?
        \item How has scheduling software fit into your prior or current work?
    \end{itemize}
    \textbf{Existing Laws and Audits/Investigations}
    \begin{itemize}
        \item There are some existing laws that affect employer scheduling and/or AI scheduling tools. Could you talk us through those laws and how they affect your work for clients?
        \item Of those laws you’ve mentioned, which would you say are...
            \begin{itemize}
                \item ... easier for you to work with? Why?
                \item ... harder for you to work with? Why?
            \end{itemize}
        \item What do you believe typically triggers a government agency investigation into a particular employer’s compliance with those laws?
        \item How does scheduling software or timekeeping records typically matter (if at all) in those kinds of investigations?
        \item Have you ever been involved in representing someone who was the subject of an agency investigation under these laws?
    \end{itemize}
    \textbf{Litigation}\\
    If possible, tell us about a legal case you worked on that involved workplace scheduling.
    \begin{itemize}
        \item What were the main allegations and legal claims?
        \item In litigating the case, what kinds of evidence did you seek from your client?
        \item How much did scheduling software matter in litigating the case? 
        \begin{itemize}
            \item In what ways did it make things easier for you?
            \item In what ways did it make things more difficult for you?
        \end{itemize}
        \item Did the plaintiff seek to examine the scheduling software in discovery or otherwise?
        \item What roles did other parties play in the case?
        \item To what extent did companies that make scheduling software become involved in the case?
        \item What was the final outcome of the case?
    \end{itemize}

\subsection{General Questions (All Stakeholders)}
\textbf{General Feelings Towards AI}\\
To begin, we’d like to learn more about your general feelings towards AI.
You may have heard about how some people are using automated software tools for work-related tasks. Some examples include assigning work schedules and managing workers’ performance.
\begin{itemize}
    \item What do you think about these kinds of (AI-assisted?) tools being used in the workplace?
    \item How do you feel about AI software that can help with or perform scheduling tasks?
    \begin{itemize}
        \item Is there anything that you would be worried about?
        \item Is there anything that you would be excited to see?
    \end{itemize}
\end{itemize}
\textbf{Value Elicitation for AI-Assisted Schedulers}
\begin{itemize}
    \item If we asked you to help us design an AI scheduler, what would you prioritize? Why are other concerns not as important?
    \item How would you compare this "ideal" AI scheduler with ones you've used or experienced in the past?
\end{itemize}
\textbf{Perceptions of Other Stakeholders’ Values}\\
Check if [X] covers the values: “accountability”, “consent”, “privacy”, “fairness”, “efficiency”, “control”.
\begin{itemize}
    \item If we asked [other stakeholder] to design the ideal AI scheduler, what do you think they'd say are its most important features?
    \begin{itemize}
        \item Why do you think [X] matters to them?
        \item How much does [X] matter to you as compared to [their most important value]?
    \end{itemize}
    \item What do you think is the greatest obstacle to having a tool that upholds [respondent’s most important value]?
    \item Are there any promising developments with algorithmic/AI scheduling that you know about that you think might help uphold these values of XYZ that you’ve identified?
\end{itemize}
\textbf{Closing Questions}
\begin{itemize}
    \item As we continue this work on designing accountable software, is there anything that we should consider?
    \item Do you know any people (regulators, worker advocates, HR/managers, or workers) that we could talk to?
\end{itemize}

\end{document}